\documentclass[prd,amsfonts,amsmath,twocolumn,superscriptaddress,nofootinbib]{revtex4}

\usepackage{color}
\usepackage{graphicx} 
\usepackage{epstopdf}%
\usepackage{placeins}

\usepackage{dblfloatfix}
\usepackage{tabularx}

\usepackage{subfloat}
\usepackage{braket}
\usepackage{mathpazo}      

\newcommand {\be} {\begin{equation}} 
\newcommand {\ba}{\begin{eqnarray}} 
\newcommand {\ee} {\end{equation}} 
\newcommand{\ea} {\end{eqnarray}}
\newcommand{\nn}{\nonumber}

\renewcommand{\epsilon}{\varepsilon}
\newcolumntype{C}[1]{>{\centering\arraybackslash}p{#1}}

\begin{document}

\title{Experimental Verification of Position-Dependent Angular-Momentum Selection Rules for Absorption of Twisted Light by a Bound Electron}

\author{Andrei Afanasev}

\affiliation{Department of Physics,
The George Washington University, Washington, DC 20052, USA}

\author{Carl E. Carlson}
\affiliation{Department of Physics, The College of William and Mary in Virginia, Williamsburg, VA 23187, USA}

\author{Christian T. Schmiegelow}
\affiliation{QUANTUM, Institut f\"ur Physik, Universit\"at Mainz, Staudingerweg 7, 55128 Mainz, Germany}
\affiliation{Departamento de Fisica, FCEyN, UBA and
IFIBA, Conicet, Pabellon 1, Ciudad Universitaria, 1428 Buenos Aires, Argentina}

\author{Jonas Schulz$^3$}

\author{Ferdinand Schmidt-Kaler$^3$}

\author{Maria Solyanik$^1$}

\date{\today}

\begin{abstract}

We analyze the multipole excitation of atoms with twisted light, i.e., by a vortex light field that carries orbital angular momentum. A single trapped $^{40}$Ca$^+$ ion serves as a localized and positioned probe of the exciting field. We drive the $S_{1/2} \to D_{5/2}$ transition and observe the relative strengths of different transitions, depending on the ion's transversal position with respect to the center of the vortex light field. On the other hand, transition amplitudes are calculated for a twisted light field in form of a Bessel beam, a Bessel-Gauss and a Gauss-Laguerre mode. Analyzing experimental obtained transition amplitudes we find agreement with the theoretical predictions at a level of better than 3\%. Finally, we propose measurement schemes with two-ion crystals to enhance the sensing accuracy of vortex modes in future experiments.
\end{abstract}

\maketitle

\section{Introduction}	\label{sec:intro}

The light with orbital angular momentum (OAM), or the twisted light has been a subject of many studies for the past 25 years. The novel features of the twisted light are due to its azimuthal phase dependence that at a quantum level results in multiple OAM-projection eigenstates that are orthogonal and therefore independently detectable, leading to the applications such as enhanced quantum communications, quantum encryption, and quantum computing. For most recent reviews of the subject, the reader is referred to Refs.\cite{Padgett2015,Franke-Arnold2017}.

In this paper we focus on the angular-momentum quantum selection rules for the excitation of quantum systems with twisted light, using atomic photoexcitation as an example. On the history of this question, it was initially shown by Babiker and collaborators \cite{Babiker2002} that in order to pass light's OAM to the {\it internal} degrees of freedom of an atom, it is required that corresponding transitions  have multipolarity higher than dipole. Direct calculations by Picon ~{\it et~al.} \cite{Picon10} of atomic photoionization demonstrated that final electrons indeed carry OAM of the incident photons. In Ref.~\cite{Afanasev:2013kaa}, it was shown that atomic photoexcitation amplitudes with the twisted light depend on atom's position through Bessel-function factors, independently of the specific atomic structure. The next step was made by authors of Ref.~\cite{Scholz2014} who derived one-to-one correspondence between twisted- and plane-wave-amplitudes for atomic photo-excitation. Based on this formalism, novel features of high-multipole transitions with twisted photons were analyzed theoretically in Ref.~\cite{Afanasev2016}, with spin-orbit effects computed in Ref.~\cite{AfanasevJOPT17}. The formalism of Ref.~\cite{Scholz2014} was extended to Laguerre-Gaussian  beams in Ref.~\cite{Peshkov17}. In other theoretical developments, the authors of Ref.~\cite{Rodrigues2015} considered excitation of Rydberg atoms with OAM beams, additional quantum selection rules with recoil effects were analyzed in Ref.~\cite{Jaregui2015}, and optical vortex interaction with multi-electron atoms was formulated in the impact-parameter space in Ref.~\cite{Kaplan15}.

Two circumstances, namely (a) the need to observe higher-multipole atomic transitions that are much weaker than dipole, and (b) high sensitivity of the transition amplitudes to atom's location within the optical vortex complicate verification of the novel quantum selection rules. The first experimental demonstration that OAM of the twisted light can be passed to the internal degrees of freedom of an atom was done recently \cite{schmiegelow2016} by measuring Rabi frequencies for $^{40}$Ca$^+$ ions placed in a Paul trap. In such a way, the ion wavepacket with an extension of $\leq$ 60~nm serves as a well-localized and positioned probe of the light field. Using an approach \cite{Schmiegelow2012} relating Rabi oscillations to dipole-like and quadrupole-like interaction operators, the authors of Ref.~\cite{schmiegelow2016} measured relative strengths of the corresponding transitions with sub-wavelength position resolution for the target $^{40}$Ca$^+$ ions. The data \cite{schmiegelow2016} appear to be sensitive to the longitudinal component of the electric field in the OAM beam \cite{Quinteiro17}.

Here we present new measurements of the complete sets of $4^2S_{1/2}\to3^2D_{5/2}$ transition amplitudes with $^{40}$Ca$^+$ ions obtained with the same apparatus as in Ref.~\cite{schmiegelow2016}. The data are presented as a function of ion's position with respect to the optical vortex center and compared with position-dependent selection rules for various OAM beam modes, namely, for Bessel, Bessel-Gauss and Laguerre-Gaussian. The results allow us to claim full understanding of the excitation strength of the atoms by the twisted light. The work is the basis of extending the studies of excitation in twisted light fields from a single ion, to the excitation of ensembled, e.g. linear trapped crystals.  We further discuss twisted light field multi-ion entanglement can be generated or, alternatively, entangled ion crystals that could be employed to analyze even with higher accuracy the polarization and vortex degrees of freedom of shaped light fields. 

The paper is organized as follows. Section II describes a theoretical formalism of quantum selection rules for twisted photoabsorption amplitudes for  various laser beam modes, predicting relative strengths of transitions into Zeeman sub-levels with given magnetic quantum numbers. Section III describes the apparatus and the experimental methods, Section IV presents comparison of the data with theory, and Section V is dedicated to summary and outlook.

\section{Twisted Light Modes and Plane Wave Factorization}	\label{sec:section 2}

\subsection{Bessel Mode}

One of the most convenient and straightforward ways to mathematically describe a beam-like behavior of EM-fields generated by lasers is by solving the scalar Helmholtz equation in cylindrical coordinates. The resulting Bessel modes were considered by Durnin \emph{et al}~\cite{Durnin:1987}, where it was also reported on first generation of Bessel beams. 

Let us briefly review the formalism described in \cite{Afanasev:2013kaa} and  consider a Bessel beam (BB) mode with total angular momentum (TAM) projection $m_{\gamma}$ which is defined with respect to the beam's propagation axis $z$. Bessel mode of frequency $\omega = |\vec{k}|$ is the family of exact normalized non-diverging solutions of the scalar wave equation in cylindrical coordinates.
\begin{equation}
\psi^{(\text{BB})}_{k_z \kappa m_\gamma}(z,\rho,t) = A \, e^{i m_\gamma \phi_\rho}  
	J_{m_\gamma}(\kappa \rho) \, e^{i(k_z z -  \omega t)}
\label{25/04/2017_4}
\end{equation}
which is mathematically defined everywhere in space, and the normalization constant is  $A=\sqrt{\kappa/2\pi}$. Here $\kappa = \sqrt{\vec{k}^2 - \vec{k}_z^2}$ is the transverse part of the wave-vector for a non-paraxial beam, $\{\rho, \phi_{\rho}, z\}$ are the cylindrical coordinates with $\vec{\rho} \perp \vec{z}$. We proceed following the notations introduced in Ref.~\cite{Afanasev:2013kaa}, and write the plane wave expansion of the Bessel mode
\begin{align}
\psi^{(\text{BB})}_{k_z \kappa m_\gamma}(\vec r,t) = e^{i(k_z z -  \omega t)}
	\int \frac{ d^2k_{\perp} }{ (2\pi)^2 }  a_{\kappa m_\gamma}(\vec k_\perp) 
		\psi^{(\text{pw})}_{\vec k}(\vec r)
\end{align}
where $\psi_{k}^{\text{pw}}(\vec{r}, t)$ are the plane wave states and $a_{k_z\kappa m_{\gamma}}(\vec{k}_{\perp})$ is the corresponding Fourier amplitude
\begin{align}
a_{\kappa m_\gamma}(\vec k_\perp) = \frac{ 2 \pi A }{ \kappa } (-i)^{m_\gamma} \,
	e^{i m_\gamma \phi_k} \, \delta( k_\perp - \kappa )
\end{align}
where $k_\perp = | \vec k_\perp |$ and $\vec k = (k_z, k_\perp, \phi_k)$.

The Bessel solutions of the wave (Helmholtz) equation for the photon vector potential can be written in the form of the superposition of plane waves with the fixed longitudinal momenta $\vec{k}_z$ and pitch angle $\theta_k = \arctan(|\vec{k}_{\perp}|/k_z)$ as follows
\begin{align}
\label{eq:vecpot}
\mathcal A^\mu_{k_z \kappa m_\gamma \Lambda}(\vec r, t) = A \, e^{i(k_z z -  \omega t)}
	\int \frac{ d{\phi_k} }{ 2\pi }  (-i)^{m_\gamma} \, e^{i m_\gamma \phi_k} \,
		e^{ i \vec k_\perp \cdot \vec \rho} \, \epsilon^\mu_{\vec k \Lambda},
\end{align}
where $\Lambda$ is helicity of a plane-wave component propagating along the direction $\vec k$.

The explicit form of the polarization state of a plane-wave photon with a wave vector $\vec k$ is
\begin{equation}
\epsilon_{\vec{k} \Lambda}^{\mu} = e^{-i\Lambda \phi_k} \cos^2 \frac{\theta_k}{2} \eta^{\mu}_{\Lambda} + e^{i \Lambda \phi_k} \sin^2 \frac{\theta_k}{2} \eta_{-\Lambda}^{\mu} + \frac{\Lambda}{ \sqrt{2}} \sin \theta_k \eta_0^{\mu}
\label{07/13/17/2}
\end{equation}
where $\{ \eta^{\mu}_{\pm \Lambda}, \eta_0^{\mu}\}$ are the polarization basis vectors
\begin{equation}
\eta^{\mu}_{\pm \Lambda} = \frac{1}{\sqrt{2}} (0, \mp \Lambda, -i, 0);\;\;\;\;\;\eta_0^{\mu} = (0,0,0,1).
\end{equation}
The local energy flux can be expressed, $e.g.$ \cite{Afanasev:2013kaa}, as a function of a pitch angle as follows
\begin{equation}
\begin{split}
f(\rho) = \cos (\theta_k) (|E|^2 + |B|^2)/4 =\;\;\;\;\;\;\;\;\;\;\;\;\;\;\;\;\;\;\;\;\;\;\;\;\;\;\;\;\;\;\\ \cos (\theta_k) \frac{A^2 \omega^2}{2} \Big\{ \cos^4 \frac{\theta_k}{2} J_{m_{\gamma} - \Lambda}^{2}(\kappa \rho)  +\sin^4 \frac{\theta_k}{2} J_{m_{\gamma} + \Lambda}^{2}(\kappa \rho) \\+ \frac{\sin^2 \theta_k}{2} J_{m_{\gamma}}^{2}(\kappa \rho) \Big\}
\end{split}
\end{equation}
where the topological effects are controlled both by the explicit $\theta_k$ - dependence in the directional cosines and the Bessel functions.

Photo-absorption of BB by the hydrogen-like atom was developed in \cite{Afanasev:2013kaa,Afanasev2014,Scholz2014, Afanasev2016}. The transition amplitude can be written in the form
\begin{align}
\mathcal M^{(\text{BB})}_{m_f m_i m_\gamma \Lambda}(b) = 
	\braket{ n_f j_f m_f | H_\text{int} | n_i j_i m_i ; \, k_z \kappa m_\gamma \Lambda ; \, b }
\end{align}
where $\{n_f,j_f,m_f\}$ and $\{n_i,j_i,m_i\}$ are respectively final and initial atomic states, and some superscripts are tacit.  Replacing the plane wave photon state by Bessel mode, and using rotation operators with the quantization axis along z-direction, we obtain the following factorization property of the twisted-wave transition amplitude, {\it c.f.} Ref.\cite{Scholz2014}:
\begin{align}
\label{eq:dsum}
\left| \mathcal M^{(\text{BB})}_{m_f m_i m_\gamma \Lambda}(b) \right|  &= 
	\bigg| J_{m_f-m_i - m_\gamma}( \kappa \rho ) \sum_{m'_f,m'_i}
	d^{j_f}_{m_f,m'_f}(\theta_k) d^{j_i}_{m_i,m'_i}(\theta_k)	\nonumber\\
&	\quad	\times	
	\mathcal M^{(\text{pw})}_{n_f j_f m'_f; \, n_i j_i m'_i; \, \Lambda} (\theta_k=0)
	\bigg|\frac{A}{2\pi}.
\end{align}
We note that since specific hydrogen-like wave functions were not used in above derivation, this factorization property also applies to other atoms and ions as long as the atomic size is much smaller than the wavelength of light. Another essential difference from Ref.~\cite{Scholz2014} is that Eq.~(\ref{eq:dsum}) applies to total angular momentum of initial and final states, including their spin. It allows consideration of both spin-dependent and spin-independent transitions for arbitrary angular-momentum eigenstates.

The general equation only requires that $H_\text{int}$ is rotation invariant.  If $H_\text{int}$ is also spin independent (as is the interaction $-(e/m) \vec p \cdot \vec A$) and if the initial state is an orbital $S$-state, the general expression can be further developed.  We can expand the final state into its orbital and spin parts, bringing in Clebsch-Gordan coefficients, and with suitable changes to the subscripts on the photoexcitation amplitude obtain for transitions from a ground state ($\l_i=0, j_i=1/2$):
\begin{align}
\label{eq:spinorbit}
&\mathcal M^{(\text{pw})}_{n_f j_f m'_f; \, n_i j_i m'_i; \, \Lambda} (\theta_k=0)	
								\nonumber\\
& \quad	= 	\left(	\begin{array}{c|cc}
			j_f	&	l_f		&	1/2	\\
			m'_f	&	l'_{fz}	&	s'_{fz}
			\end{array}		\right)
\mathcal M^{(\text{pw})}_{n_f l_f l'_{fz}, \, 1/2 s'_{fz}; \, n_i, 1/2, m'_i; \, \Lambda} (\theta_k=0)
								\nonumber\\
& \quad	= 	\left(	\begin{array}{c|cc}
			j_f	&	l_f		&	1/2	\\
			m'_f	&	\Lambda	&	m'_i
			\end{array}		\right)
\mathcal M^{(\text{pw})}_{n_f l_f \Lambda,  \, n_i , \, \Lambda} (\theta_k=0)	,
\end{align}
where we remember that $H_\text{int}$ is spin independent for this development and the last matrix element is calculated using only the orbital parts of the electron states.

With suitable manipulation, the overall matrix element can be given as a product with no sums, and for a transition to a fine structure state with fixed $l_f$ as well as fixed $j_f$,
 \begin{align}
 \label{eq:bbfact}
\left| \mathcal M^{(\text{BB})}_{m_f m_i m_\gamma \Lambda} (b) \right|  & = 
	\bigg| J_{m_f-m_i - m_\gamma}( \kappa \rho )
		\left(	\begin{array}{c|cc}
			j_f	&	l_f		&	1/2	\\
			m_f	&	m_f-m_i	&	m_i
			\end{array}		\right)
	 	\nonumber\\
&	\times
	d^{l_f}_{m_f - m_i, \Lambda}(\theta_k)	
	\mathcal M^{(\text{pw})}_{n_f l_f \Lambda,  \, n_i , \, \Lambda} (\theta_k=0)
	\bigg|\frac{A}{2\pi}
\end{align}

Two main effects related to the topology of the incoming photon state should be noticed: rotational transformation described by the Wigner {\it d}-function and topological phase factor $J_{m_f - m_{\gamma}} (\kappa b)$. These two novel factors in the absorption amplitude modify the angular momentum selection rules for BB vs the plane-wave case. 
In the electron-photon interaction we neglect effects of electron spin that are in general suppressed for atomic photo-excitation if electric multipoles are allowed. For this reason we can replace the difference $m_f-m_i=m_{lf}-m_{li}$, where $m_{li} (m_{lf})$ are OAM projections of initial (final) electron states. We emphasize that the above formalism of Eq.(\ref{eq:dsum}) automatically includes electron-spin-dependent interaction, while the next step, i.e. separation into the orbital and spin part of the electron wave function (\ref{eq:spinorbit}) implies that the electron spin remains intact during photo-excitation.

\subsection{Bessel-Gauss Mode} \label{sec:section 3}

Bessel modes accurately describe the observed behavior of EM-fields at the beam center. However, for the peripheral behavior the diverging nature of this solution of the Maxwell equations becomes non-negligible. A convenient generalization of the fundamental Bessel mode, Bessel-Gauss mode (BG), was first considered by Sheppard and Wilson \cite{sheppard1978gaussian}. It belongs to the family of Helmholtz-Gauss beams and satisfies the paraxial wave equation. Its characteristic behavior mimics BB in vicinity to the quantization axis, while secondary maxima get strongly suppressed by the Gaussian factor
\begin{align}
\psi^{(\text{BG})}_{\kappa m_\gamma}(z,\vec \rho,t) = A \, e^{i m_\gamma \phi_\rho}  
	J_{m_\gamma}(\kappa \rho) \, e^{i(k_z z -  \omega t)} \, e^{-\rho^2/w_0^2}
\end{align}
where  $A$ is the overall constant coming from the Fresnel expansion (e.g., see \cite{bagini1996generalized}). Other parameters are defined identically to the conventions of the Gaussian and Bessel modes: $\text{w}_0$ is the waist of the beam and $\theta_k$ is the pitch angle).  

After taking 2D Fourier transform one can obtain the following form for the plane wave expansion of the Bessel-Gauss mode:
\begin{align}
\psi^{(\text{BG})}_{\kappa m_\gamma}(\vec r,t) = e^{i(k_z z -  \omega t)}
	\int \frac{ d^2k_\perp }{ (2\pi)^2 }  a^{(\text{BG})}_{\kappa m_\gamma}(\vec k_\perp) 
		e^{-i k_\perp \cdot \vec \rho}
\end{align}
where the integral is taken over the entire reciprocal space, similar to the angular spectrum representation technique \cite{novotny2012principles}, and the contribution coming from evanescent waves ($k_{\perp} \in [k, \infty)$) is negligible. The corresponding Fourier kernel is 
\begin{equation}
\begin{split}
a_{\kappa m_{\gamma}}^{(\text{BG})}(k_{\perp}) = A \pi i^{m_{\gamma}} e^{im_{\gamma}\phi_k} \text{w}_0^2 \;\;\;\;\;\;\;\;\;\;\;\;\;\;\;\;\;\;\;\;\;\;\;\;\;\;\; \\ \exp{\Big[-\frac{\kappa^2+k_{\perp}^2}{4} \text{w}_0^2 \Big]} I_{m_{\gamma}}\Big( \frac{\kappa \text{w}_0^2}{2} k_{\perp} \Big)
\end{split}
\end{equation}
The function $I_{m_{\gamma}}(z) = i^{m_{\gamma}}J_{m_{\gamma}}(iz)$ is the modified Bessel function. Applying the formalism laid out in, e.g. \cite{Afanasev2016, Scholz2014}, one can obtain for electron-spin-independent part of the transition amplitude:
\begin{align}
& \left| \mathcal M^{(\text{BG})}_{m_{lf} m_{li}=0; m_\gamma \Lambda}(b) \right|
	= \bigg|  \frac{ A w_0^2 }{ 4\pi } 	\,
	\mathcal M^{(\text{pw})}_{n_f l_f \Lambda, \, n_i \Lambda}(\theta_k = 0)	\,
	e^{ - \kappa^2 w_0^2 / 4 }										\nonumber\\
&\times		\int k_\perp dk_\perp		\,  d^{l_f}_{m_{lf} \Lambda}(\theta_{k})	  \,
	J_{m_\gamma-m_{lf}}(k_\perp b) \,  
	I_{m_\gamma}\big( \frac{1}{2} w_0^2 \kappa k_\perp \big)  \,
	e^{ - w_0^2  k_\perp^2  / 4 }		\bigg|.
\label{07/13/2017/4}	
\end{align}
This integral can be calculated, e.g. \cite{gradshteyn2014table} (6.633 1), involving an infinite sum over hypergeometric functions. To include the effect of electron spin in the atomic fine structure, Clebsch-Gordan coefficient factors have to be applied as in Eq.~(\ref{eq:bbfact}) above.

When the parameter $w_0$ is large compared with other dimensional quantities such as the wavelength, we can evaluate the Wigner function at the pitch angle $\theta_\kappa$ and take it out of the integral. Further, we can approximate the modified Bessel function by its asymptotic value and evaluate the integral explicitly, obtaining
\begin{align}
& \left| \mathcal M^{(\text{BG})}_{m_{lf}, m_{li}=0, m_\gamma }(b) \right|
	= 	e^{-b^2/w_0^2 } 			\nonumber\\
&\qquad	\times 	\bigg|	\frac{ A  }{ 2\pi } 		\,
	J_{m_\gamma-m_{lf}}(\kappa b)   \,  
	d^{l_f}_{m_{lf} \Lambda}(\theta_k)	  \,
	\mathcal M^{(\text{pw})}_{n_f l_f \Lambda, \, n_i\Lambda}(\theta_k = 0)	\bigg|	\,,
\end{align}
which is a Gaussian factor times the result Eq.(\ref{eq:bbfact}) for a pure non-Gaussian Bessel beam, with relevant Clebsch-Gordan coefficients implied. That the Gaussian modification of the starting beam profile feeds through in such a simple way to the photoexcitation amplitude works only if $w_0$ is large.  For parameters of interest to us, $w_0$ is large enough and the two photoexcitation amplitude expressions give nearly identical numerical results.  

\subsection{Laguerre-Gaussian Mode} \label{sec:section 4}

Laguerre - Gaussian (LG) mode plays a fundamental role in photonics, laser optics and resonators \cite{siegman1986university, teich1991fundamentals}. It belongs to the family of Gaussian solutions the scalar paraxial equation. The spatial amplitude dependence expressed by the equation
\begin{equation}
\begin{split}
\psi_{LG}(\vec{\rho}, t=0; z) = \Big( \frac{\rho \sqrt{2}}{\text{w}(z)} \Big)^{|\ell_{\gamma}|} L_p^{|\ell_{\gamma}|} \Big( \frac{2\rho^2}{\text{w}^2(z)} \Big) e^{-\frac{\rho^2}{\text{w}^2(z)}} \\ e^{-i\frac{k\rho^2 z}{2(z^2+z_R^2)}}e^{i |\ell_{\gamma}| \phi_\rho+i k_z z} \frac{c}{\sqrt{1+z^2/z_R^2}} e^{i\phi_{G}}+ c.c.
\label{25/04/2017_3}
\end{split}
\end{equation}
Here $\ell_{\gamma}$ is the beam vorticity factor that coincides with its OAM projection in paraxial approximation; $\text{w}_z = \text{w}_0 \sqrt{1+(z/z_R)^2}$ is its spotsize; $z_R= k\text{w}_0/2$ is the Rayleigh range; $\phi_G=\arctan(z/z_R)$ is the Gouy phase of the LG mode. The associate Laguerre polynomial is given, as can be found elsewhere, e.g. \cite{abramowitz1964handbook,gradshteyn2014table}
\begin{equation}
L_p^{|\ell_{\gamma}|}\Big( \frac{2\rho^2}{\text{w}^2} \Big) = \sum_{j=0}^p (-1)^j \frac{(|\ell_{\gamma}| + p)!}{(p-j)!(|\ell_{\gamma}| + j)! j!} \Big( \frac{2\rho^2}{\text{w}^2} \Big)^j
\end{equation}
where $p$ is the number of radial nodes ($p+1$ concentric circles). Bessel function, being a complete set of orthogonal functions, can be used as an expansion basis. Here we will consider the mode in focus $z=0$ and perform the Hankel transform given as
\begin{gather}
f(x) = \int_{D_{\xi}} \xi F(\xi) J_{\gamma}(\xi x) d\xi = F^{-1} [F(\xi);x]\\
F(\xi) = \int_{D_{\xi}} x f(x) J_{\gamma} (\xi x)dx = F[f(x);\xi] \label{25/04/2017_2}
\end{gather}
to obtain the expression for the LG-mode expanded in Bessel modes \eqref{25/04/2017_4},
where $\xi \in D_{\xi}$ and $x \in D_x$. We assume that the transformation kernels are symmetric, such as
\begin{equation}
K(x;\xi) = \xi J_{\gamma} (\xi x);\;\;\;\;K(\xi; x) = x J_{\gamma} (\xi x).
\end{equation}
After applying \eqref{25/04/2017_2} to \eqref{25/04/2017_3} and with the help of the transform \cite{magnus1954tables}
\begin{equation}
\mathcal{H}_{\nu} [x^{\nu} e^{-\frac{1}{4}px^2};\xi] = \frac{2^{\nu+1} \xi^{\nu}}{p^{\nu+1}} e^{-\xi^2/p}
\end{equation}
we get the following expression for the scalar LG mode
\begin{equation}
\begin{split}
\psi_{LG}(\vec{\rho}, \phi_r, z) = \sum_{j=0}^{p} \mathcal{B}_{pj}^{|\ell_{\gamma}|} \;\;\;\;\;\;\;\;\;\;\;\;\;\;\;\;\;\;\;\;\;\;\;\;\;\;\; \\ \sqrt{2\pi}\int_0^{\infty} k_{\perp}^{2j+|\ell_{\gamma}|+\frac{1}{2}} e^{-k_{\perp}^2\text{w}_0^2/4} \psi_{k_{\perp} |\ell_{\gamma}| k_z}^{BB}\Big(\vec{\rho}\Big) d k_{\perp},
\label{25/04/2017_5}
\end{split}
\end{equation}
where the BB state is given as in \eqref{25/04/2017_4} and $\mathcal{B}_{pj}^{|\ell_{\gamma}|}$ is the expansion coefficient defined as
\begin{equation}
\mathcal{B}_{pj}^{|\ell_{\gamma}|} = \frac{(-1)^j(|\ell_{\gamma}| + p)!}{(p-j)!(|\ell_{\gamma}| +j)!j!} \Big( \frac{\text{w}_0}{\sqrt{2}} \Big)^{2j+|\ell_{\gamma}| + 2}
\end{equation}
As a result, the vector solution of the paraxial wave equation for LG mode can be expressed as follows
\begin{equation}
\begin{split}
\mathcal{A}_{LG}^{\mu} =  \sum_{j=0}^{p} \mathcal{B}_{pj}^{|\ell_{\gamma}|} \sqrt{2\pi} \\ \int_0^{\infty} d k_{\perp} \; k_{\perp}^{2j+|\ell_{\gamma}| +\frac{1}{2}} e^{-k_{\perp}^2\text{w}_0^2/4} \mathcal{A}_{k_{\perp} m_{\gamma} k_z \Lambda}^{\mu} (\vec{r},t)
\label{11/21/2016/1}
\end{split}
\end{equation}
with $\mathcal{A}_{k_{\perp} m_{\gamma} k_z \Lambda}^{\mu} (\vec{r},t)$ defined as in \eqref{eq:vecpot}. The polarization basis \eqref{07/13/17/2} was taken in its paraxial form
\begin{equation}
\vec{\epsilon}_{k\Lambda} = 
- \frac{\Lambda}{\sqrt{2}}  \{1,i\Lambda, 0\} e^{-i\Lambda \phi_k}
\label{01/11/2017_1}
\end{equation}

Making use of the approach developed for BB's earlier, we arrive at the following expression
\begin{equation}
\begin{split}
|M_{LG}(\vec{\rho}, \phi_r, z)| = \Big|\sqrt{\frac{\kappa}{2\pi}} \sum_{j=0}^p \mathcal{B}_{pj}^{|\ell_{\gamma}|}  \sum_{m'_{lf} m'_{li}} d_{m_{lf}m'_{lf}}^{\ell_f} (\theta_k) d_{m_{li}m'_{li}}^{\ell_i} (\theta_k) \\ \mathcal{M}^{\text{(pw)}}_{n_f \ell_f \Lambda n_i \Lambda}(0) \int_0^{\infty}dk_{\perp}\; k_{\perp}^{2j+|\ell_{\gamma}|+1} e^{-k_{\perp}^2\text{w}_0^2/4} J_{m_{\gamma}+m_{li}-m_{lf}} (k_{\perp} b) \Big|
\label{11/28/2016/2}
\end{split}
\end{equation}
where similarly to \eqref{eq:bbfact} and \eqref{07/13/2017/4} factorization is possible. The integral on the right-hand side can be calculated analytically as in \cite{gradshteyn2014table}, (6.643 4) with the variable substitution $x=k_{\perp}^2$. This leads to the following representation of the transition amplitude for $\ell_i = m_{li} = 0$ - S-state, where the original LG-beam structure is apparent
\begin{equation}
\begin{split}
|M_{LG}(\vec{\rho}, \phi_r, z)| = \Big|\sqrt{\frac{\kappa}{2\pi}} \sum_{j=0}^p \tilde{\mathcal{B}}_{pj}^{|\ell_{\gamma}|}  \sum_{m'_{lf} m'_{li}} d_{m_{lf}m'_{lf}}^{\ell_f} (\theta_k) d_{m_{li}m'_{li}}^{\ell_i} (\theta_k) \\ \mathcal{M}_{n_f \ell_f \Lambda n_i \Lambda}^{(\text{pw})} (\theta_k=0) \Big( \frac{b}{2} \Big)^{|\eta|} e^{-b^2/\text{w}_0^2} \mathcal{L}_{j+\xi}^{|\eta|} \Big( \frac{b^2}{\text{w}_0^2} \Big) \Big|
\label{07/13/2017/1}
\end{split}
\end{equation}
The parameters are $\xi = 0.5(|\ell_{\gamma}|-|\eta|)$ and $\eta = m_{\gamma} + m_{li} - m_{lf}$. The new coefficient is
\begin{equation}
\tilde{\mathcal{B}}_{pj}^{|\ell_{\gamma}|} = \frac{\text{sgn}(\eta)^{\eta}(-1)^j(|\ell_{\gamma}| + p)! \lfloor(j+\xi)\rceil!}{2^{-j-(|\ell_{\gamma}|+|\eta|)/2}(p-j)!(|\ell_{\gamma}| +j)!j!} \Big( \frac{2}{\text{w}_0} \Big)^{|\eta|}
\end{equation}
where $\text{sgn}(\bullet)$ denotes the signum of the number, $\lfloor \bullet \rceil!$ is the Roman factorial \cite{roman1992logarithmic} and $\mathcal{L}_{n}^{|\nu|} (\bullet)$ is the extended Laguerre polynomial
\begin{equation}
\mathcal{L}_{n}^{|\nu|} (x)= \frac{\lfloor (n+|\nu|) \rceil !}{\lfloor n \rceil! \lfloor |\nu| \rceil!} \;_1F_1(-n, \alpha+1, x)
\end{equation}

\section{Description of the Experiment} \label{sec:experiment}

We now describe the experimental procedure to determine the position dependent selection rules. We position the single ion in the vortex light field and determine the its variation of excitation strength for various magnetic transitions between Zeeman sublevels of the $S_{1/2} \to D_{5/2} $ transition. 

A single trapped 
$^{40}$Ca$^+$ ion is trapped and Doppler cooled in a segmented Paul trap to a thermal state with wave packet size of about 60 nm. The ion position along one of the transverse axis of the probe beam is controlled by applying programmable voltages on the trap electrodes. 
This allows positioning the ion with sub nanometer precision along one beam axis.

Experiments follow a sequence of (i) Doppler cooling, (ii) followed by optical pumping... (iii) Then the ion is shifted ... Each experiment starts with Doppler cooling followed by optical pumping at a fixed position. The ion was then shifted to its probe position along the beam axis where the quadrupole beam was turned on for a given amount of time. Then the ion was brought again to its initial position where state readout was performed by state selective fluorescence.

As a probe beam we use a Ti:Sa
laser tuned to the $ 4S_{1/2} $-$ 3D_{5/2} $ transition near 729 nm.
Its frequency is stabilized to better than 100 Hz by
locking to a high-finesse ULE reference cavity (Stable-Lasers inc.) cavity with finesse close to 200,000. 
Buy use of an acousto-optic modulator (AOM) in a double pass configuration the laser can be switched and tuned to the different Zeeman transitions spanning a range of 30 Mhz.
This allows us to probe all the 
transitions
$| 4^2S_{1/2},mJ = \pm \frac{1}{2} \rangle \leftrightarrow  
 | 3^2D_{5/2}, mJ = \pm \frac{1}{2}, \pm \frac{3}{2}, \pm \frac{5}{2} \rangle$, which are Zeeman split by an external field of 13 mT. After passing the AOM, the light beam, the beam is coupled into a polarization maintaining single-mode fiber for spatial filtering and to deliver it near the experiment where it is out-coupled into free space.

We use two different spatial distributions for the beam: a plain Gaussian beam and vortex with chirality one. For the Gaussian beam we take the shape as filtered by the single mode fiber. To produce the vortex beams with chirality $\pm 1$ we 
additionally place a holographic fork shaped phase plate 
in the beam path.  A full description of the apparatus is given in~\cite{schmiegelow2016} and shown in Fig.\ref{fig:experiment} here.

The polarization of the beam is then set by a half wave-plate on a motorized mount and a combination of quarter-half-quarter wave-plates and dielectric mirrors. 
The combination of wave-plates and mirrors are set so that by rotating the first motorized wave-plate, the polarization of the beam before focusing onto the ion can be chosen to be either circular left or right. The need of the extra wave plates is to compensate for the polarization changes on the subsequent dielectric mirrors. Calibration of polarizations was done with a polarization analyzer (Schaefter-Kirchhoff SK010PA-NIR) and a metallic pick up mirror after the last dichroic mirror. Polarizations were set to a Stokes parameter $S_3 = 1$ with a accuracy of 1\%.  However because of slight misalignment of the calibration procedure we expect the actual polarizations to be correct between 1-3\%. This number varies in different experimental runs because re-calibration of the polarization was redone periodically.

The probe beam was focused on the ion 
by the use of a 50 mm achromatic lens with a 67 mm focal length and 50 mm diameter objective, which allowed focusing to a waist of 2.7~$\mu$m.

Each profile scan was done by probing the ion at different positions with a fixed interrogation time. This time was chosen so that it would never exceed the pi time at any point of the beam. The measured excitation probability $P$ is related to the Rabi frequency $\Omega$ by 
$P = (1 - \cos(\Omega t))/2$. By inverting this formula we obtain the Rabi frequencies from the measured probabilities. For each position the experiment was repeated 100 times. The reported value is the mean with an error given by the Clopper-Pearson confidence interval at 1 sigma.

Additionally, due to the frequency-dependent diffraction efficiency of the acousto-optic modulator used to tune the laser, the laser power was different for each measured transition. Additionally, different powers were chosen in some cases to provide a better dynamic range. To account for these changes we re-scaled the obtained Rabi frequencies with the square root of the optical power used in each case.

Finally we note that the magnetic field was set by a combination of magnetic coils which were set to nullify the earth and room magnetic fields and to set a field along the beam axis of 13 mT. The sense of this field could be inverted to change the chirality of the beam's angular momentum with respect to the ion�s. This option was chosen, rather than 
flipping the phase plate, because it provided a more reproducible way of shifting between beam types. The reason for this is that this procedure did not involve complete re-alignment of the beam path and its focus as the rotation of the phase plate does.

\section{Comparison of Data with Theory Predictions}

Let us compare the experimental data with theoretical predictions. To adjust theory parameters, we first use the data on normalized Rabi frequencies for two transitions, with $\Delta m=-2,-1$ (where $\Delta m$ is the magnetic quantum number change in the atomic state) shown in Fig.~{\ref{fig:BB_BG_m12}  caused by OAM photons with wavelength $\lambda$=729~nm, total angular momentum projection $m_\gamma=-2$ and spin helicity $\Lambda=-1$, i.e. with photon's OAM aligned with its spin.

We start with BB mode that has only two independent parameters: (a)~an overall normalization and (b)~the pitch angle $\theta_k$. Fixing the normalization to reproduce $m_\gamma=-2$ amplitude at zero impact parameter, and choosing $\theta_k$=0.095~rad to reproduce positions of first minima for $\Delta m=-2,-1$ transitions, we see from Fig.{\ref{fig:BB_BG_m12}a (green dashed and blue solid lines) that while BB mode reproduces the data well in the central region of the beam at $b\leq 3 \mu$m, it overshoots the data at larger impact parameters. Introducing BG mode into the comparison, we see that the data are well reproduced with a choice of an additional parameter $w_0^{BG}=10~\mu$m=13.7$\lambda$ (Fig.~{\ref{fig:BB_BG_m12}a, red dotted and black short-dashed lines). BG mode behaves nearly identical to BB in vicinity to the optical vortex center for most transitions, however in case with BG partial amplitudes get suppressed on the beam's periphery, which better reflects the physical behavior of the laser beam.

Theoretical description with LG mode of a given order $p$ (Fig.\ref{fig:BB_BG_m12}b) requires three independent parameters: 1)~overall normalization factor; 2) waist $w_0$ that fixes the position of the minima and controls overall height of the maxima; and 3) pitch angle $\theta_k$ that controls relative height of the maxima. 
When comparing the behavior of the LG beam to the data we notice that for the case of $p=0$ both the first node in beam intensity and the second maxima are either absent or largely suppressed compared to the experimental observation, while choosing $p=1$ for LG mode alone overshoots the data at large impact parameters.
We found that the theory describes the data the best if the mixture of LG-modes with the orders $p=0$ and $p=1$ is considered Fig.\ref{fig:BB_BG_m12}b. The waists for these two contributions were adjusted independently by parameters $w_0$ and $w_1$ accordingly. Treating the relative ratio of $p$=1 and $p$=0 LG modes as an independent parameter, our LG model has five parameters in total. Using the same pitch angle as in BG mode $\theta_k$=0.095~rad, we find the optimal values of other parameters as follows: $w_0=4.0~\mu$m, $w_1=6.5~\mu$m, $p$=1 to $p$=0 mode ratio (by amplitude) =0.43.

Keeping all the above parameters unchanged, and choosing the opposite spin $\Lambda=1$, we can predict the amplitudes for $m_\gamma=0$ , when photons spin and OAM are anti-aligned. Corresponding intensity profiles of BB, BG and LG beams with the above choice of parameters are shown in Fig.~\ref{fig:flux} as a function of the radial distance to the optical vortex center. One can see that different theory models of the beam give similar intensity profiles  in the central region and start to noticeably differ at $b\geq 5-6~\mu$m. 


{\it Without} further adjusting the theory parameters, full data sets with corresponding calculations are given as grids in Figs.~\ref{fig:fullmBG} (BG model) and \ref{fig:fullmLG} (LG model). The transitions all have $j_f = 5/2$ and $l_f=2$, and $s_{iz} = -1/2$.  Each plot has the impact parameter on the horizontal axis and the reduced Rabi frequency, or a number proportional to the transition amplitude magnitude, on the vertical axis.

For BB and BG beams, Bessel functions determine where the zeros of the amplitude lie, which includes determining if the amplitude is zero at zero impact parameter  $b$.  The cases with finite amplitude at zero impact parameter have $m_f = m_i + m_\gamma$ and are highlighted in red on the data plots.  On the other hand, the Bessel functions all have about the same peak magnitude for all index values, so the Bessel functions are not decisive for setting the relative scale of the different data sets.

Additional factors come from the Clebsch-Gordan coefficients.  Figs.~\ref{fig:fullmBG},\ref{fig:fullmLG} have $m_i = -1/2$ for all data sets, and for the grid in this figure, each element of a given column has the same $m_f$, with the values from left to right given as $m_f= -5/2, -3/2, -1/2,\, 1/2,\, 3/2$.  The corresponding Clebsch-Gordan coefficients for the columns are
\begin{align}
\label{eq:cg1}
\sqrt{ \frac{5}{5} }, \quad \sqrt{ \frac{4}{5} }, \quad \sqrt{ \frac{3}{5} }, \quad \sqrt{ \frac{2}{5} }, \quad \sqrt{ \frac{1}{5} }	\,,
\end{align}
in order from left to right.  

More decisive for the size of the predicted amplitude is the Wigner $d$-functions.  For small angles, they are proportional to powers of the angle, as
\begin{align}
\label{eq:tm}
d^{j}_{m_f - m_i, \Lambda}(\theta_k)   \propto   ( \theta_k )^{| m_f - m_i - \Lambda | }  .
\end{align}
Since $\theta_k \approx 0.1$~rad for the data in the paper, the value of the exponent is decisive in setting the overall scale of each amplitude.  

The first three rows of Figs. ~\ref{fig:fullmBG},\ref{fig:fullmLG} all have $\Lambda = -1$ and the last three rows all have 
$\Lambda = 1$.   Hence as we go across any of the first three rows from left to right, we can expect the $d$-functions to give factors proportional to
\begin{align}
\label{eq:tp}
( \theta_k )^1, \quad ( \theta_k )^0, \quad ( \theta_k )^1, \quad ( \theta_k )^2, \quad ( \theta_k )^3.
\end{align}
The effects of these factors upon the normalization of the amplitudes is easily seen in the labeling of the vertical axes of the respective figures.  The corresponding factors for the last three rows are
\begin{align}
( \theta_k )^3, \quad ( \theta_k )^2, \quad ( \theta_k )^1, \quad ( \theta_k )^0, \quad ( \theta_k )^1,
\end{align}
and the effects of these factors are equally easy to see.

For corresponding data sets with $m_i = +1/2$, Figs.~\ref{fig:fullpBG},\ref{fig:fullpLG}, the $m_f$ values are again the same for each element of a given column, and are arranged as $m_f = -~3/2,-~1/2,\,1/2,\,3/2,\,5/2$.   The Clebsch-Gordan coefficients are in order just the reverse of Eq.~\eqref{eq:cg1}.   The factors of $\theta_k$ are, however, the same as in Eqs.~\eqref{eq:tm} and~\eqref{eq:tp}.  (The value of $s_{iz}$ has changed by one unit, but the $m_f$ labeling of the columns is also offset by one unit.)

Using the theoretical model, we can evaluate the radius of {\it prenumbra} introduced in Ref.\cite{schmiegelow2016} that we can define as an impact parameter $b$ for which $\Delta m=-2$ transition with $m_\gamma=-2$ (forbidden for plane-wave photons) equals $\Delta m=-1$ (allowed for plane waves). For BB modes the estimate is most straightforward,
\begin{align}
\label{eq:prenum}
s_z&=-\frac{1}{2}; \ \ J_0(\kappa b) d_{21}^{(2)}(\theta_k)=\sqrt{\frac{4}{5}}J_1(\kappa b)d_{11}^{(2)}(\theta_k) \\ \nonumber
s_z&=\frac{1}{2}; \ \ \ \sqrt{\frac{1}{5}}J_0(\kappa b) d_{21}^{(2)}(\theta_k)=\sqrt{\frac{2}{5}}J_1(\kappa b)d_{11}^{(2)}(\theta_k).
\end{align}
Expanding above expressions for small $\theta_k$, we obtain for {\it prenumbra} radius $b=\lambda \frac{\sqrt{5}}{2\pi}=0.26\mu$m for $s_z=-1/2$ and 
$b=\frac{\lambda}{\pi \sqrt{2}}=0.16\mu$m for $s_z=1/2$, that can be checked against Fig.\ref{fig:BB_BG_m12}. It is about twice as large for the initial electron spin aligned with photon's TAM projection $m_\gamma$.

Let us compare the peak values of non-vanishing amplitudes at the optical vortex center ($b=0$) -- indicated by red plots in Figs.~4-7 with theory predictions, see Tables~\ref{table1},\ref{table2}, with data taken from Ref.\cite{schmiegelow2016} (Supplementary Material).  They appear to be in good agreement. The relative peak values at zero impact parameter are determined only by Wigner functions for BB and (factorized) BB modes. For example, for $m_\gamma=-2,\Lambda=-1$ vs  for $m_\gamma=0,\Lambda=1$ we compare Wigner $d$-functions $d^{(2)}_{-2-1}(\theta_k)$ ($\Delta m=-2$) and $d^{(2)}_{01}(\theta_k)$ (for $\Delta m=0$): the latter is larger by a factor $\sqrt{3/2}$ in a small-angle limit, predicting that the ratio of squares of the corresponding Rabi frequencies should equal 1.5. Forming an electron spin-averaged sum of the squared Rabi frequencies from the Tables~1,2, we find the experimental value of the ratio =1.48(8) to be a good match. This is also an experimental evidence that the twisted light should develop circular dichroism in the high-multipole absorption by unpolarized atomic target, the effect predicted in Ref.\cite{AfanasevJOPT17}. A different approach was used in Ref.~\cite{Quinteiro17} to predict ratios of $b=0$ Rabi frequencies, where the role of longitudinal component of the electric was analyzed; this approach is consistent with the one presented here.

\begin{table}[]
\centering
\caption{Measured Rabi frequencies from Ref.[1] in units of kHz/$\mu$W taken at zero impact parameter b = 0 and compared to theoretical predictions with overall normalization fixed to $\Delta m_{\gamma} = -2$ transition and $s_z=-1/2$ initial state.}
\label{table1}
\begin{tabular}{|p{2cm}||C{1.5cm}|C{1.5cm}|C{1.5cm}|C{1.5cm}|C{1.5cm}}
\hline
$S_z=-1/2$ & BB & BG & LG & Data \\ \hline \hline
$\Delta m=-2$ & 2.92 & 2.92 & 2.92 & 2.92 (8) \\ \hline
$\Delta m=-1$ & 27.1 & 29.7 & 21.7 & 31.21 (87) \\ \hline
$\Delta m=0$ & 2.76 & 3.11 & 2.76 & 2.78 (8) \\ \hline
$\Delta m=0$ & 2.76 & 3.11 & 2.76 & 2.78 (7) \\ \hline
$\Delta m=1$ & 19.2 & 21. & 15.3 & 19.22 (62) \\ \hline
$\Delta m=2$ & 1.3 & 1.31 & 1.31 & 1.26 (4) \\ \hline
\end{tabular}
\end{table}

\begin{table}[]
\centering
\caption{Same as Table \ref{table1} with overall normalization fixed to $\Delta m_{\gamma} = 2$ transition and  $s_z=1/2$ initial state.}
\label{table2}
\begin{tabular}{|p{2cm}||C{1.5cm}|C{1.5cm}|C{1.5cm}|C{1.5cm}|C{1.5cm}}
\hline
$S_z=1/2$ & BB & BG & LG & Data \\ \hline \hline
$\Delta m=-2$ & 1.24 & 1.24 & 1.24 & 1.33 (4) \\ \hline
$\Delta m=-1$ & 18.2 & 19.9 & 14.5 & 23.89 (66) \\ \hline
$\Delta m=0$ & 2.62 & 2.95 & 2.62 & 2.87 (8) \\ \hline
$\Delta m=0$ & 2.62 & 2.95 & 2.62 & 2.61 (8) \\ \hline
$\Delta m=1$ & 25.7 & 28.2 & 20.6 & 34.08 (92) \\ \hline
$\Delta m=2$ & 2.77 & 2.77 & 2.77 & 2.77 (8) \\ \hline
\end{tabular}
\end{table}

The theory calculations initially assumed that the laser beam is fully circularly polarized, i.e. $\Lambda=1$ (LCP) or -1 (RCP). Comparison with data suggested, however, that the beams are slightly elliptic, with deviations from fully circular polarization at 1\% or less, that is well within the accuracy of measured polarization, as outlined in Sec. \ref{sec:experiment}. Adding the amplitudes of opposite helicity (but the same OAM) with appropriate weight bring the theory and data within agreement. Comparing the green solid plots with black dashed plots in Figs.~\ref{fig:fullmBG} to \ref{fig:fullpLG}, one can see that the transition amplitudes most affected by this small ellipticity are the amplitudes that have large-strength counterparts with opposite circular polarization. Namely, for  $\Lambda=-1$ the amplitudes significantly modified by the opposite-sign polarization admixture are for $\Delta m=1,2$, and vise versa for the opposite $\Lambda$.

We also present data for the topological charge $\overline m_\gamma \equiv m_\gamma - \Lambda = \pm 2$, with the corresponding theory predictions, in Figs.~\ref{fig:OAM2_BG} and~\ref{fig:OAM2_LG}. In the paraxial limit, $\overline m_\gamma$ would be the orbital angular momentum.  The previous figures all had $\left| \overline m_\gamma \right| \le 1$. We find here that the nonzero transition at the vortex center for spin and OAM \textit{anti-aligned} is indeed nonzero, indicating that photon's OAM fully reversed the sign of the magnetic quantum number compared to the plane-wave case. However an admixture of the opposite-helicity photon state at 10\% (by amplitude) obscures this effect away from the vortex center. 

Finally, let us discuss possible azimuthal dependence of the atomic transitions. Inspecting $\phi_b$-dependence of the factorized transition amplitude for OAM light that is due to the fact that both photons and electrons are eigenstates of angular momentum projection on the propagation direction, we see an overall phase factor $e^{i (m_\gamma + m_i - m_f ) \phi_b}$, as in Eq.~(\ref{eq:BBFacPhase}) in Appendix. Hence in the measurement with a fixed value of $m_\gamma$ the overall phase does not affect observables. However, for OAM light the light beam can be a coherent superposition of the states with different values of $m_\gamma$, as in linearly polarized beams, for example. Again defining $\overline m_\gamma=m_\gamma-\Lambda$, we can derive transition amplitudes with linearly polarized beams by adding the amplitudes with opposite values of $\Lambda$, while keeping $\overline m_\gamma$ fixed. The result reveals an azimuthal dependence that is small for all amplitudes, except $\Delta m =  0$, for which azimuthal variation is significant as shown in Fig.~\ref{fig:azimuthal}.  The upper plots in Fig.~\ref{fig:azimuthal} show the excitation amplitude magnitudes for four different azimuthal angles, as indicated in the caption, and the lower plots give the azimuthal dependence as contour plots in the $x$-$y$ plane, with lighter colors indicating large amplitude and darker colors indicating small amplitude.

\section{Summary and Outlook}						\label{sec:summary}

We have presented extensive data on the photoexcitation of atomic states by twisted photons, along with a theoretical study of the selection rules and impact parameter dependence pertinent to this process.   The theory and the data are in good agreement.

All data are for $4 ^2S_{1/2}$ to $3^2D_{5/2}$ transitions in once ionized $^{40}$Ca.  Transitions with the target atom both on and off the photon vortex axis were measured for all possible $D_{5/2}$ final states, for both possible polarizations of the initial state, and for a variety of angular momenta of the twisted photon states.  In all, there are 60 data sets, presented in Figs.~4 to 7.

When the atom is on the vortex axis, there is a selection rule that the angular momentum of the photon must all be absorbed into the final electronic state.   This can give high magnetic quantum number final states, and is in marked contrast to what is possible with plane wave photons. The  selection rule was first observed empirically in~\cite{schmiegelow2016}, and is seen clearly in the present data.  The most relevant cases are highlighted with red data points in Figs. 4-7.

When the target atom is away from the vortex center, the data is well predicted by theory using either Bessel-Gauss or Laguerre-Gauss descriptions of the twisted photon beam.

The theory for the excitation amplitudes depends on four parameters.  Three of them are the overall normalization, the spatial width of the beam, and the pitch angle.  The fourth parameter measures the small amplitude of opposite helicity photons, in a beam nominally made from photons of a single helicity $\Lambda$.   All data sets are, excepting a few cases where the data is too sparse to make a judgement, in good agreement with predictions based on these few parameters.

The experiments and theory on atomic  photoexcitation verify and enhance our understanding of twisted photon states. They may also eventually become valuable diagnostic tools.  It has already been noted the the on-axis selection rules are a way to determine a beam's vorticity.  Further, since other parameters can be determined from fitting, with the demonstrated success of such fits,  measurements like those shown here may additionally become a tool to deduce other beam characteristics such as width, pitch angle, and helicity composition.  

Usual atomic spectra are dominated by electric dipole transitions. However, as transition rates increase with the degree of ionization, the high-multipole transitions become important in highly charged ions. The longest lifetimes are commonly observed in moderately charged ions \cite{trabert2000atomic}. 
These properties are widely used for diagnostics of astrophysical and laboratory plasmas, making highly charged ions good candidates for atomic clocks. 
For the case of OAM photons, we expect to see modified transition rates with the characteristic impact parameter dependence. The effect is going to be the highest for the atoms located in the central region of the laser spot, relaxing to the plane-wave-like behavior on the beam periphery. This is going to be especially evident for the transitions different in order and multipolarity, but compatible in rates. This is a common situation in plasma spectroscopy, {\it e.g.} \cite{safronova2006relativistic}.

The storage and recall of photon states are crucial for realizing OAM quantum memory \cite{nicolas2014quantum,Franke-Arnold2017}. The presented results should be instrumental for developing quantum computing with OAM light, and provide experimentally verified foundation for atomic spectroscopy with twisted light.

Our experiments and the comparison with the theoretical models have proven that a {\em single ion} serves as a high-precision and well-localized probe of light fields with complex vortex and polarization structures. We plan to extend the method of probing such fields with a {\em pair of two ions}. Here, the inter-ion distance allows for an accurate ruler of the length scale and the entire crystal is scanned in position through the beam profile. We plan to observe the ion crystals' fluorescence, but now with a CCD camera that allows for parallel and independent readout of both ions. Small differences of excitation would be detected with much higher accuracy. Ultimately, {\it quantum entangled pairs} of ions in a specific sensor Bell state of Zeeman sublevels $ \{ m=+1/2, -1/2 \}$ in the S$_{1/2}$ ground state would be generated. The ion crystal in this state would be exposed to the vortex field and a different AC-Stark shift would be induced for both ions. This results in phase shift difference, and consequently, a Bell state $\Psi_+=\sqrt{2} (|+1/2, -1/2 \rangle + |+1/2, -1/2 \rangle)$ will undergo a parity oscillation between $\Psi_+$ and $\Psi_-$, which is finally detected by a quantum state analysis. Recent work with Bell states have demonstrated the advantages of quantum entanglement for magnetic field difference measurements~\cite{Ruster17}, in quite similar way this technique would lead to orders-of-magnitude improvements for sensing structured light fields.

\section*{Acknowledgements}

Work of AA and MS was partially supported by U.S. Army Research Office Grant No.~W911NF-16-1-0452 and Gus Weiss Endowment of The George Washington University.
FSK acknowledges partial support by the Office of the Director of National Intelligence via the grant W911NF-16-1-0070. CEC thanks the National Science Foundation for support under Grant PHY-1516509.
\appendix

\section{Wigner rotations of the states} 

Here we give some detail regarding how twisted photon matrix elements are related to plane wave matrix elements, where the electron's spin is included in the eigenstates of total angular momentum, c.f. Ref.~\cite{Scholz2014} where only OAM degrees of freedom were considered.  

In general, the twisted photon matrix element can be given in terms of plane wave photon matrix elements using the Fourier decomposition,
\begin{align}
\label{eq:finalme}
\mathcal M^{(m_\gamma)}_{m_f m_i \Lambda}(\vec b) &= \braket{n_f  j_f m_f  | H_1 | 
	n_i j_i m_i ;\  k_\perp k_z m_\gamma \Lambda ; \vec b }			\nn\\
	&=  A \int \frac{ d\phi_k }{ 2\pi } \ i^{-m_\gamma} 
	e^{i m_\gamma \phi_k - i \vec k_\perp \cdot \vec b }			\nn\\
&\quad \times	\braket{n_f j_f m_f  | H_1 | 
	n_i j_i m_i  ;\  k_\perp k_z \phi_k \Lambda }	.
\end{align}
The last matrix element has atomic states quantized along the $z$-axis, but the photon momentum not in the $z$-direction.  We isolate the plane wave matrix element as
\begin{align}
\mathcal M^{\text{(pw)}}_{m_f m_i \Lambda}(\theta_k,\phi_k) &=
	\braket{n_f j_f m_f  | H_1 | n_i j_i m_i  ;\  k_\perp k_z \phi_k \Lambda }	.
\end{align}
The technique for evaluating this matrix element is to rotate the states so that the photon's momentum is along the $z$-direction, and re-expressing the rotated the atomic states in terms of states quantized along the $z$-direction, using known properties of rotations.

The photon state is, with the phase convention of~\cite{Jentschura:2010ap},
\begin{align}
\ket{k_\perp k_z \phi_k \Lambda} = R(\phi_k,\theta_k) \ket{ k, \Lambda} 
	= R_z(\phi_k) R_y(\theta_k) \ket{ k, \Lambda}	,
\end{align}
where the last ket represents a state moving in the $z$-direction.  The Hamiltonian is rotation invariant, so that
\begin{align}
\mathcal M^{\text{(pw)}}_{m_f m_i \Lambda}(\theta_k,\phi_k) &=
	\braket{n_f j_f m_f  | \, R \, H_1 | \left(R^\dagger (n_i j_i m_i) \right) ;\ k \Lambda }	.
\end{align}
The rotated atomic states are related to states quantized along the $z$-axis by the Wigner rotation matrices,  leading to
\begin{align}
\label{eq:pw1}
\mathcal M^{\text{(pw)}}_{m_f m_i \Lambda}(\theta_k,\phi_k)
 &=	e^{-i(m_f-m_i)\phi_k}		\sum_{m'_i}
	d^{j_f}_{m_f,m'_f = m'_i+\Lambda}(\theta_k) 
			\nn\\
&\hskip -0 em	\quad \times	d^{j_i}_{m_i,m'_i}(\theta_k)
	\mathcal M^{\text{(pw)}}_{m'_f=m'_i+\Lambda, m'_i, \Lambda}(0,0)	,
\end{align}
where
\begin{align}
\mathcal M^{\text{(pw)}}_{m'_f, m'_i, \Lambda}(0,0) =
	\braket{n_f j_f m'_f  |  H_1 | n_i j_i m'_i ;\ k \Lambda }			.
\end{align}
Given that the momentum and state quantization are now all in the $z$-direction, it follows that $m'_f = m'_i + \Lambda$.  

One can further develop the result by expanding the states in an $LS$ basis.  For atomic applications, only the electric part of the electromagnetic interaction Hamiltonian is needed, and the electric part is spin independent.  For simplicity and for direct use in this paper, we will just  the case where the initial state is an orbital $S$-state, or initial atomic orbital angular momentum $l_i =0$.

The total initial angular momentum is just the initial spin, $j_i = s_i$, and the total projection is just the spin projection, $m_i = s_{iz}$.  The plane wave matrix element for arbitrary photon direction becomes,
\begin{align}
\mathcal M^{\text{(pw)}}_{m_f m_i \Lambda}(\theta_k,\phi_k) &=
	\sum_{m'_f l_{fz} \sigma m'_i}		\braket{j_f m_f | R | j_f m'_f}
		\left(		\begin{array}{c|cc}
				j_f		&	l_f		&	s_f	\\
				m'_f	&	l_{fz}	&	\sigma
				\end{array}				\right)					\nn\\
&\hskip -4.5 em	\times 
		\braket{n_f  l_f l_{fz} s_f \sigma |  H_1 | n_i \, 0 0 \, s_i m'_i ; \ k \Lambda }		\ 
		\braket{ s_i m'_i | R^\dagger | s_i m_i }	.
\end{align}
From the spin independence of $H_1$, we obtain  $s_f = s_i$, with both being $s_e = 1/2$ when only one electron is under consideration, and the spin projections are the same.  Thus,
\begin{align}
\label{eq:pw2}
\mathcal M^{\text{(pw)}}_{m_f m_i \Lambda}(\theta_k,\phi_k) 
	&=  e^{-i (m_f - m_i ) \phi_k}		\sum_{m'_f, \sigma}	
		d^{j_f}_{m_f m'_f}(\theta_k)	\,
			d^{s_e}_{m_i m'_i}(\theta_k)					\nn\\
&\hskip 0.7 em	\times
		\left(		\begin{array}{c|cc}
				j_f				&	l_f		&	s_e	\\
				m'_f	&	\Lambda	&	m'_i
				\end{array}				\right)
		\mathcal M^{\text{(pw)}}_{l_{fz}= \Lambda, 0, \Lambda}(0,0)	,
\end{align}
where
\begin{align}
\mathcal M^{\text{(pw)}}_{l_{fz}, 0, \Lambda}(0,0) = 
	\braket{n_f  l_f l_{fz} |  H_1 | n_i \, 0 0 ;\ k \Lambda }
\end{align}
is calculated with only the orbital wave functions.

The result is not yet at its simplest.  The sum can be eliminated.  One way to do this is to work with the definition of the Wigner functions and with the $LS$ expansion to show
\begin{align}
d^{j_f}_{m_f m'_f}(\theta_k) &= 
		\left(		\begin{array}{c|cc}
				j_f				&	l_f		&	s_f	\\
				m_f	&	l_z	&	s_z
				\end{array}				\right)
		\left(		\begin{array}{c|cc}
				j_f				&	l_f		&	s_f	\\
				m'_f	&	l'_z	&	s'_z
				\end{array}				\right)
		d^{l_f}_{l_z l'_z}(\theta_k)
		d^{s_e}_{s_z s'_z}(\theta_k),
\end{align}
and by substitution, summing on Clebsch-Gordan coefficients, and multiplying Wigner functions, obtain the identity
\begin{align}
&\sum_{m'_f, m'_i}	
		d^{j_f}_{m_f m'_f}(\theta_k)	\,
			d^{s_e}_{m_i m'_i}(\theta_k)
		\left(		\begin{array}{c|cc}
				j_f				&	l_f		&	s_e	\\
				m'_f	&	\Lambda	&	m'_i
				\end{array}				\right)			\nn\\
&\hskip 4 em	=	\left(		\begin{array}{c|cc}
				j_f	&	l_f			&	s_f	\\
				m_f	&	m_f - m_i	&	m_i
				\end{array}				\right)
		d^{l_f}_{{m_f-m_i}, \Lambda}(\theta_k)		.
\end{align}
Hence,
\begin{align}
\label{eq:pw3}
\mathcal M^{\text{(pw)}}_{m_f m_i \Lambda}(\theta_k,\phi_k) 	&= e^{-i(m_f-m_i)\phi_k}
		\left(		\begin{array}{c|cc}
				j_f	&	l_f			&	s_f	\\
				m_f	&	m_f - m_i	&	m_i
				\end{array}				\right)
														\nn\\[1 ex]
&\hskip 0.7 em	\times
	d^{l_f}_{m_f-m_i, \Lambda}(\theta_k)	\, 
	\mathcal M^{\text{(pw)}}_{\Lambda, 0, \Lambda}(0,0) .
\end{align}

The same form can alternatively be obtained beginning with the plane wave matrix element for arbitrary photon direction, and doing the $LS$ expansions and taking the spin matrix elements before doing any rotations.

Finally, to finish the calculation of $\mathcal M^{(m_\gamma)}_{m_f m_i \Lambda}(\vec b)$ as given in Eq.~\eqref{eq:finalme}, substitute the plane wave matrix element, either Eq.~\eqref{eq:pw1} or \eqref{eq:pw2} or \eqref{eq:pw3}, into Eq.~\eqref{eq:finalme}, and do the $\phi_k$ integral.  This gives a Bessel function.  Using the last, no sum, result as an example, 
\begin{align}
\label{eq:BBFacPhase}
\mathcal M^{(m_\gamma)}_{m_f m_i \Lambda}(\vec b) &=A \, 
	i^{m_i -m_f } \ 
	e^{i (m_\gamma + m_i - m_f ) \phi_b} 	\, 	J_{m_f - m_i - m_\gamma}(k_\perp b)  \,
										\nn\\
&\hskip -4. em	\times	d^{l_f}_{m_f-m_i, \Lambda}(\theta_k)
			\left(		\begin{array}{c|cc}
				j_f	&	l_f			&	s_f	\\
				m_f	&	m_f - m_i	&	m_i
				\end{array}				\right)
	\mathcal M^{\text{(pw)}}_{l_{fz}= \Lambda, 0, \Lambda}(0,0) .
\end{align}

This is the form we use in the body of the paper for the Bessel and Bessel-Gauss mode evaluations.  Notice that only one $\mathcal M^{\text{(pw)}}(0,0)$ is needed for all the transition possibilities studied here.  This can be interpreted as requiring only one overall normalization constant to fit all the data discussed in the text of this article. 
The application of above formalism to $4^2S_{1/2}$ to $3^2D_{5/2}$ transitions is discussed in Section IV.

\bibliography{TwistedPhoton}

\begin{thebibliography}{31}
\expandafter\ifx\csname natexlab\endcsname\relax\def\natexlab#1{#1}\fi
\expandafter\ifx\csname bibnamefont\endcsname\relax
  \def\bibnamefont#1{#1}\fi
\expandafter\ifx\csname bibfnamefont\endcsname\relax
  \def\bibfnamefont#1{#1}\fi
\expandafter\ifx\csname citenamefont\endcsname\relax
  \def\citenamefont#1{#1}\fi
\expandafter\ifx\csname url\endcsname\relax
  \def\url#1{\texttt{#1}}\fi
\expandafter\ifx\csname urlprefix\endcsname\relax\def\urlprefix{URL }\fi
\providecommand{\bibinfo}[2]{#2}
\providecommand{\eprint}[2][]{\url{#2}}

\bibitem[{\citenamefont{Wisniewski-Barker and Padgett}(2015)}]{Padgett2015}
\bibinfo{author}{\bibfnamefont{E.}~\bibnamefont{Wisniewski-Barker}}
  \bibnamefont{and} \bibinfo{author}{\bibfnamefont{M.}~\bibnamefont{Padgett}},
  \bibinfo{journal}{Photonics: Scientific Foundations, Technology and
  Applications,} \textbf{\bibinfo{volume}{1}}, \bibinfo{pages}{321}
  (\bibinfo{year}{2015}).

\bibitem[{\citenamefont{Franke-Arnold}(2017)}]{Franke-Arnold2017}
\bibinfo{author}{\bibfnamefont{S.}~\bibnamefont{Franke-Arnold}},
  \bibinfo{journal}{Philosophical Transactions of the Royal Society of London
  A: Mathematical, Physical and Engineering Sciences}
  \textbf{\bibinfo{volume}{375}} (\bibinfo{year}{2017}), ISSN
  \bibinfo{issn}{1364-503X},
  \urlprefix\url{http://rsta.royalsocietypublishing.org/content/375/2087/20150435}.

\bibitem[{\citenamefont{Babiker et~al.}(2002)\citenamefont{Babiker, Bennett,
  Andrews, and D\'avila~Romero}}]{Babiker2002}
\bibinfo{author}{\bibfnamefont{M.}~\bibnamefont{Babiker}},
  \bibinfo{author}{\bibfnamefont{C.~R.} \bibnamefont{Bennett}},
  \bibinfo{author}{\bibfnamefont{D.~L.} \bibnamefont{Andrews}},
  \bibnamefont{and} \bibinfo{author}{\bibfnamefont{L.~C.}
  \bibnamefont{D\'avila~Romero}}, \bibinfo{journal}{Phys. Rev. Lett.}
  \textbf{\bibinfo{volume}{89}}, \bibinfo{pages}{143601}
  (\bibinfo{year}{2002}),
  \urlprefix\url{http://link.aps.org/doi/10.1103/PhysRevLett.89.143601}.

\bibitem[{\citenamefont{Pic\'on et~al.}(2010)\citenamefont{Pic\'on, Mompart,
  de~Aldana, Plaja, Calvo, and Roso}}]{Picon10}
\bibinfo{author}{\bibfnamefont{A.}~\bibnamefont{Pic\'on}},
  \bibinfo{author}{\bibfnamefont{J.}~\bibnamefont{Mompart}},
  \bibinfo{author}{\bibfnamefont{J.~R.~V.} \bibnamefont{de~Aldana}},
  \bibinfo{author}{\bibfnamefont{L.}~\bibnamefont{Plaja}},
  \bibinfo{author}{\bibfnamefont{G.~F.} \bibnamefont{Calvo}}, \bibnamefont{and}
  \bibinfo{author}{\bibfnamefont{L.}~\bibnamefont{Roso}},
  \bibinfo{journal}{Optics Express} \textbf{\bibinfo{volume}{18}},
  \bibinfo{pages}{3660} (\bibinfo{year}{2010}), \eprint{1002.1318}.

\bibitem[{\citenamefont{Afanasev et~al.}(2013)\citenamefont{Afanasev, Carlson,
  and Mukherjee}}]{Afanasev:2013kaa}
\bibinfo{author}{\bibfnamefont{A.}~\bibnamefont{Afanasev}},
  \bibinfo{author}{\bibfnamefont{C.~E.} \bibnamefont{Carlson}},
  \bibnamefont{and}
  \bibinfo{author}{\bibfnamefont{A.}~\bibnamefont{Mukherjee}},
  \bibinfo{journal}{Phys. Rev. A,} \textbf{\bibinfo{volume}{88}},
  \bibinfo{pages}{033841} (\bibinfo{year}{2013}).

\bibitem[{\citenamefont{Scholz-Marggraf
  et~al.}(2014)\citenamefont{Scholz-Marggraf, Fritzsche, Serbo, Afanasev, and
  Surzhykov}}]{Scholz2014}
\bibinfo{author}{\bibfnamefont{H.~M.} \bibnamefont{Scholz-Marggraf}},
  \bibinfo{author}{\bibfnamefont{S.}~\bibnamefont{Fritzsche}},
  \bibinfo{author}{\bibfnamefont{V.~G.} \bibnamefont{Serbo}},
  \bibinfo{author}{\bibfnamefont{A.}~\bibnamefont{Afanasev}}, \bibnamefont{and}
  \bibinfo{author}{\bibfnamefont{A.}~\bibnamefont{Surzhykov}},
  \bibinfo{journal}{Phys. Rev. A} \textbf{\bibinfo{volume}{90}},
  \bibinfo{pages}{013425} (\bibinfo{year}{2014}),
  \urlprefix\url{http://link.aps.org/doi/10.1103/PhysRevA.90.013425}.

\bibitem[{\citenamefont{Afanasev et~al.}(2016)\citenamefont{Afanasev, Carlson,
  and Mukherjee}}]{Afanasev2016}
\bibinfo{author}{\bibfnamefont{A.}~\bibnamefont{Afanasev}},
  \bibinfo{author}{\bibfnamefont{C.~E.} \bibnamefont{Carlson}},
  \bibnamefont{and}
  \bibinfo{author}{\bibfnamefont{A.}~\bibnamefont{Mukherjee}},
  \bibinfo{journal}{Journal of Optics} \textbf{\bibinfo{volume}{18}},
  \bibinfo{pages}{074013} (\bibinfo{year}{2016}),
  \urlprefix\url{http://stacks.iop.org/2040-8986/18/i=7/a=074013}.

\bibitem[{\citenamefont{Afanasev et~al.}(2017)\citenamefont{Afanasev, Carlson,
  and Solyanik}}]{AfanasevJOPT17}
\bibinfo{author}{\bibfnamefont{A.}~\bibnamefont{Afanasev}},
  \bibinfo{author}{\bibfnamefont{C.~E.} \bibnamefont{Carlson}},
  \bibnamefont{and} \bibinfo{author}{\bibfnamefont{M.}~\bibnamefont{Solyanik}},
  \bibinfo{journal}{Journal of Optics}  (\bibinfo{year}{2017}),
  \urlprefix\url{http://iopscience.iop.org/10.1088/2040-8986/aa82c3}.

\bibitem[{\citenamefont{Peshkov et~al.}(2017)\citenamefont{Peshkov, Seipt,
  Surzhykov, and Fritzsche}}]{Peshkov17}
\bibinfo{author}{\bibfnamefont{A.~A.} \bibnamefont{Peshkov}},
  \bibinfo{author}{\bibfnamefont{D.}~\bibnamefont{Seipt}},
  \bibinfo{author}{\bibfnamefont{A.}~\bibnamefont{Surzhykov}},
  \bibnamefont{and}
  \bibinfo{author}{\bibfnamefont{S.}~\bibnamefont{Fritzsche}},
  \bibinfo{journal}{Phys. Rev. A} \textbf{\bibinfo{volume}{96}},
  \bibinfo{pages}{023407} (\bibinfo{year}{2017}),
  \urlprefix\url{https://link.aps.org/doi/10.1103/PhysRevA.96.023407}.

\bibitem[{\citenamefont{{Rodrigues} et~al.}(2015)\citenamefont{{Rodrigues},
  {Marcassa}, and {Mendon{\c c}a}}}]{Rodrigues2015}
\bibinfo{author}{\bibfnamefont{J.~D.} \bibnamefont{{Rodrigues}}},
  \bibinfo{author}{\bibfnamefont{L.~G.} \bibnamefont{{Marcassa}}},
  \bibnamefont{and} \bibinfo{author}{\bibfnamefont{J.~T.}
  \bibnamefont{{Mendon{\c c}a}}}, \bibinfo{journal}{ArXiv e-prints}
  (\bibinfo{year}{2015}), \eprint{1512.05933}.

\bibitem[{\citenamefont{J\'auregui}(2015)}]{Jaregui2015}
\bibinfo{author}{\bibfnamefont{R.}~\bibnamefont{J\'auregui}},
  \bibinfo{journal}{Phys. Rev. A} \textbf{\bibinfo{volume}{91}},
  \bibinfo{pages}{043842} (\bibinfo{year}{2015}),
  \urlprefix\url{http://link.aps.org/doi/10.1103/PhysRevA.91.043842}.

\bibitem[{\citenamefont{Kaplan and McGuire}(2015)}]{Kaplan15}
\bibinfo{author}{\bibfnamefont{L.}~\bibnamefont{Kaplan}} \bibnamefont{and}
  \bibinfo{author}{\bibfnamefont{J.~H.} \bibnamefont{McGuire}},
  \bibinfo{journal}{Phys. Rev. A} \textbf{\bibinfo{volume}{92}},
  \bibinfo{pages}{032702} (\bibinfo{year}{2015}),
  \urlprefix\url{http://link.aps.org/doi/10.1103/PhysRevA.92.032702}.

\bibitem[{\citenamefont{Schmiegelow et~al.}(2016)\citenamefont{Schmiegelow,
  Schulz, Kaufmann, Ruster, Poschinger, and Schmidt-Kaler}}]{schmiegelow2016}
\bibinfo{author}{\bibfnamefont{C.~T.} \bibnamefont{Schmiegelow}},
  \bibinfo{author}{\bibfnamefont{J.}~\bibnamefont{Schulz}},
  \bibinfo{author}{\bibfnamefont{H.}~\bibnamefont{Kaufmann}},
  \bibinfo{author}{\bibfnamefont{T.}~\bibnamefont{Ruster}},
  \bibinfo{author}{\bibfnamefont{U.~G.} \bibnamefont{Poschinger}},
  \bibnamefont{and}
  \bibinfo{author}{\bibfnamefont{F.}~\bibnamefont{Schmidt-Kaler}},
  \bibinfo{journal}{Nature Communications} \textbf{\bibinfo{volume}{7}},
  \bibinfo{pages}{12998} (\bibinfo{year}{2016}),
  \urlprefix\url{http://dx.doi.org/10.1038/ncomms12998}.

\bibitem[{\citenamefont{Schmiegelow and Schmidt-Kaler}(2012)}]{Schmiegelow2012}
\bibinfo{author}{\bibfnamefont{C.}~\bibnamefont{Schmiegelow}} \bibnamefont{and}
  \bibinfo{author}{\bibfnamefont{F.}~\bibnamefont{Schmidt-Kaler}},
  \bibinfo{journal}{Eur. Phys. J. D} \textbf{\bibinfo{volume}{66}},
  \bibinfo{pages}{157} (\bibinfo{year}{2012}),
  \urlprefix\url{http://dx.doi.org/10.1140/epjd/e2012-20730-4}.

\bibitem[{\citenamefont{{Quinteiro} et~al.}(2017)\citenamefont{{Quinteiro},
  {Schmidt-Kaler}, and {Schmiegelow}}}]{Quinteiro17}
\bibinfo{author}{\bibfnamefont{G.~F.} \bibnamefont{{Quinteiro}}},
  \bibinfo{author}{\bibfnamefont{F.}~\bibnamefont{{Schmidt-Kaler}}},
  \bibnamefont{and} \bibinfo{author}{\bibfnamefont{C.~T.}
  \bibnamefont{{Schmiegelow}}}, \bibinfo{journal}{ArXiv e-prints}
  (\bibinfo{year}{2017}), \eprint{1707.04776}.

\bibitem[{\citenamefont{Durnin}(1987)}]{Durnin:1987}
\bibinfo{author}{\bibfnamefont{J.}~\bibnamefont{Durnin}}, \bibinfo{journal}{J.
  Opt. Soc. Am. A} \textbf{\bibinfo{volume}{4}}, \bibinfo{pages}{651}
  (\bibinfo{year}{1987}).

\bibitem[{\citenamefont{Afanasev et~al.}(2014)\citenamefont{Afanasev, Carlson,
  and Mukherjee}}]{Afanasev2014}
\bibinfo{author}{\bibfnamefont{A.}~\bibnamefont{Afanasev}},
  \bibinfo{author}{\bibfnamefont{C.~E.} \bibnamefont{Carlson}},
  \bibnamefont{and}
  \bibinfo{author}{\bibfnamefont{A.}~\bibnamefont{Mukherjee}},
  \bibinfo{journal}{J. Opt. Soc. Am. B} \textbf{\bibinfo{volume}{31}},
  \bibinfo{pages}{2721} (\bibinfo{year}{2014}).

\bibitem[{\citenamefont{Sheppard and Wilson}(1978)}]{sheppard1978gaussian}
\bibinfo{author}{\bibfnamefont{C.}~\bibnamefont{Sheppard}} \bibnamefont{and}
  \bibinfo{author}{\bibfnamefont{T.}~\bibnamefont{Wilson}},
  \bibinfo{journal}{IEE Journal on Microwaves, Optics and Acoustics}
  \textbf{\bibinfo{volume}{2}}, \bibinfo{pages}{105} (\bibinfo{year}{1978}).

\bibitem[{\citenamefont{Bagini et~al.}(1996)\citenamefont{Bagini, Frezza,
  Santarsiero, Schettini, and Spagnolo}}]{bagini1996generalized}
\bibinfo{author}{\bibfnamefont{V.}~\bibnamefont{Bagini}},
  \bibinfo{author}{\bibfnamefont{F.}~\bibnamefont{Frezza}},
  \bibinfo{author}{\bibfnamefont{M.}~\bibnamefont{Santarsiero}},
  \bibinfo{author}{\bibfnamefont{G.}~\bibnamefont{Schettini}},
  \bibnamefont{and} \bibinfo{author}{\bibfnamefont{G.~S.}
  \bibnamefont{Spagnolo}}, \bibinfo{journal}{Journal of Modern Optics}
  \textbf{\bibinfo{volume}{43}}, \bibinfo{pages}{1155} (\bibinfo{year}{1996}).

\bibitem[{\citenamefont{Novotny and Hecht}(2012)}]{novotny2012principles}
\bibinfo{author}{\bibfnamefont{L.}~\bibnamefont{Novotny}} \bibnamefont{and}
  \bibinfo{author}{\bibfnamefont{B.}~\bibnamefont{Hecht}},
  \emph{\bibinfo{title}{Principles of Nano-Optics}}
  (\bibinfo{publisher}{Cambridge University Press}, \bibinfo{year}{2012}).

\bibitem[{\citenamefont{Gradshteyn and Ryzhik}(2014)}]{gradshteyn2014table}
\bibinfo{author}{\bibfnamefont{I.~S.} \bibnamefont{Gradshteyn}}
  \bibnamefont{and} \bibinfo{author}{\bibfnamefont{I.~M.}
  \bibnamefont{Ryzhik}}, \emph{\bibinfo{title}{Table of Integrals, Series, and
  Products}} (\bibinfo{publisher}{Academic Press}, \bibinfo{year}{2014}).

\bibitem[{\citenamefont{Siegman}(1986)}]{siegman1986university}
\bibinfo{author}{\bibfnamefont{A.~E.} \bibnamefont{Siegman}},
  \bibinfo{journal}{Mill Valley, CA} \textbf{\bibinfo{volume}{37}}
  (\bibinfo{year}{1986}).

\bibitem[{\citenamefont{Teich and Saleh}(1991)}]{teich1991fundamentals}
\bibinfo{author}{\bibfnamefont{M.~C.} \bibnamefont{Teich}} \bibnamefont{and}
  \bibinfo{author}{\bibfnamefont{B.}~\bibnamefont{Saleh}},
  \bibinfo{journal}{Canada, Wiley Interscience} \textbf{\bibinfo{volume}{3}}
  (\bibinfo{year}{1991}).

\bibitem[{\citenamefont{Abramowitz and Stegun}(1964)}]{abramowitz1964handbook}
\bibinfo{author}{\bibfnamefont{M.}~\bibnamefont{Abramowitz}} \bibnamefont{and}
  \bibinfo{author}{\bibfnamefont{I.~A.} \bibnamefont{Stegun}},
  \emph{\bibinfo{title}{Handbook of mathematical functions: with formulas,
  graphs, and mathematical tables}}, vol.~\bibinfo{volume}{55}
  (\bibinfo{publisher}{Courier Corporation}, \bibinfo{year}{1964}).

\bibitem[{\citenamefont{Magnus et~al.}(1954)\citenamefont{Magnus, Bateman,
  Erd{\'e}lyi, Oberhettinger, and Tricomi}}]{magnus1954tables}
\bibinfo{author}{\bibfnamefont{W.}~\bibnamefont{Magnus}},
  \bibinfo{author}{\bibfnamefont{H.}~\bibnamefont{Bateman}},
  \bibinfo{author}{\bibfnamefont{A.}~\bibnamefont{Erd{\'e}lyi}},
  \bibinfo{author}{\bibfnamefont{F.}~\bibnamefont{Oberhettinger}},
  \bibnamefont{and} \bibinfo{author}{\bibfnamefont{F.}~\bibnamefont{Tricomi}},
  \emph{\bibinfo{title}{Tables of integral transforms}}
  (\bibinfo{publisher}{McGraw-Hill}, \bibinfo{year}{1954}).

\bibitem[{\citenamefont{Roman}(1992)}]{roman1992logarithmic}
\bibinfo{author}{\bibfnamefont{S.}~\bibnamefont{Roman}}, \bibinfo{journal}{The
  American Mathematical Monthly} \textbf{\bibinfo{volume}{99}},
  \bibinfo{pages}{641} (\bibinfo{year}{1992}).

\bibitem[{\citenamefont{Tr{\"a}bert}(2000)}]{trabert2000atomic}
\bibinfo{author}{\bibfnamefont{E.}~\bibnamefont{Tr{\"a}bert}},
  \bibinfo{journal}{Physica Scripta} \textbf{\bibinfo{volume}{61}},
  \bibinfo{pages}{257} (\bibinfo{year}{2000}).

\bibitem[{\citenamefont{Safronova et~al.}(2006)\citenamefont{Safronova,
  Safronova, Hamasha, and Beiersdorfer}}]{safronova2006relativistic}
\bibinfo{author}{\bibfnamefont{U.}~\bibnamefont{Safronova}},
  \bibinfo{author}{\bibfnamefont{A.}~\bibnamefont{Safronova}},
  \bibinfo{author}{\bibfnamefont{S.}~\bibnamefont{Hamasha}}, \bibnamefont{and}
  \bibinfo{author}{\bibfnamefont{P.}~\bibnamefont{Beiersdorfer}},
  \bibinfo{journal}{Atomic Data and Nuclear Data Tables}
  \textbf{\bibinfo{volume}{92}}, \bibinfo{pages}{47} (\bibinfo{year}{2006}).

\bibitem[{\citenamefont{Nicolas et~al.}(2014)\citenamefont{Nicolas, Veissier,
  Giner, Giacobino, Maxein, and Laurat}}]{nicolas2014quantum}
\bibinfo{author}{\bibfnamefont{A.}~\bibnamefont{Nicolas}},
  \bibinfo{author}{\bibfnamefont{L.}~\bibnamefont{Veissier}},
  \bibinfo{author}{\bibfnamefont{L.}~\bibnamefont{Giner}},
  \bibinfo{author}{\bibfnamefont{E.}~\bibnamefont{Giacobino}},
  \bibinfo{author}{\bibfnamefont{D.}~\bibnamefont{Maxein}}, \bibnamefont{and}
  \bibinfo{author}{\bibfnamefont{J.}~\bibnamefont{Laurat}},
  \bibinfo{journal}{Nature Photonics} \textbf{\bibinfo{volume}{8}},
  \bibinfo{pages}{234} (\bibinfo{year}{2014}).

\bibitem[{\citenamefont{Ruster et~al.}(2017)\citenamefont{Ruster, Kaufmann,
  Luda, Kaushal, Schmiegelow, Schmidt-Kaler, and Poschinger}}]{Ruster17}
\bibinfo{author}{\bibfnamefont{T.}~\bibnamefont{Ruster}},
  \bibinfo{author}{\bibfnamefont{H.}~\bibnamefont{Kaufmann}},
  \bibinfo{author}{\bibfnamefont{M.}~\bibnamefont{Luda}},
  \bibinfo{author}{\bibfnamefont{V.}~\bibnamefont{Kaushal}},
  \bibinfo{author}{\bibfnamefont{C.}~\bibnamefont{Schmiegelow}},
  \bibinfo{author}{\bibfnamefont{F.}~\bibnamefont{Schmidt-Kaler}},
  \bibnamefont{and}
  \bibinfo{author}{\bibfnamefont{U.}~\bibnamefont{Poschinger}},
  \bibinfo{journal}{arXive 1704.01793}  (\bibinfo{year}{2017}).

\bibitem[{\citenamefont{Jentschura and Serbo}(2011)}]{Jentschura:2010ap}
\bibinfo{author}{\bibfnamefont{U.}~\bibnamefont{Jentschura}} \bibnamefont{and}
  \bibinfo{author}{\bibfnamefont{V.}~\bibnamefont{Serbo}},
  \bibinfo{journal}{Phys.Rev.Lett.} \textbf{\bibinfo{volume}{103}},
  \bibinfo{pages}{013001} (\bibinfo{year}{2011}), \eprint{eprint
  arXiv:1008.4788}.

\end{thebibliography}

\onecolumngrid

\begin{figure}[h]
\centering
\includegraphics[scale=0.95]{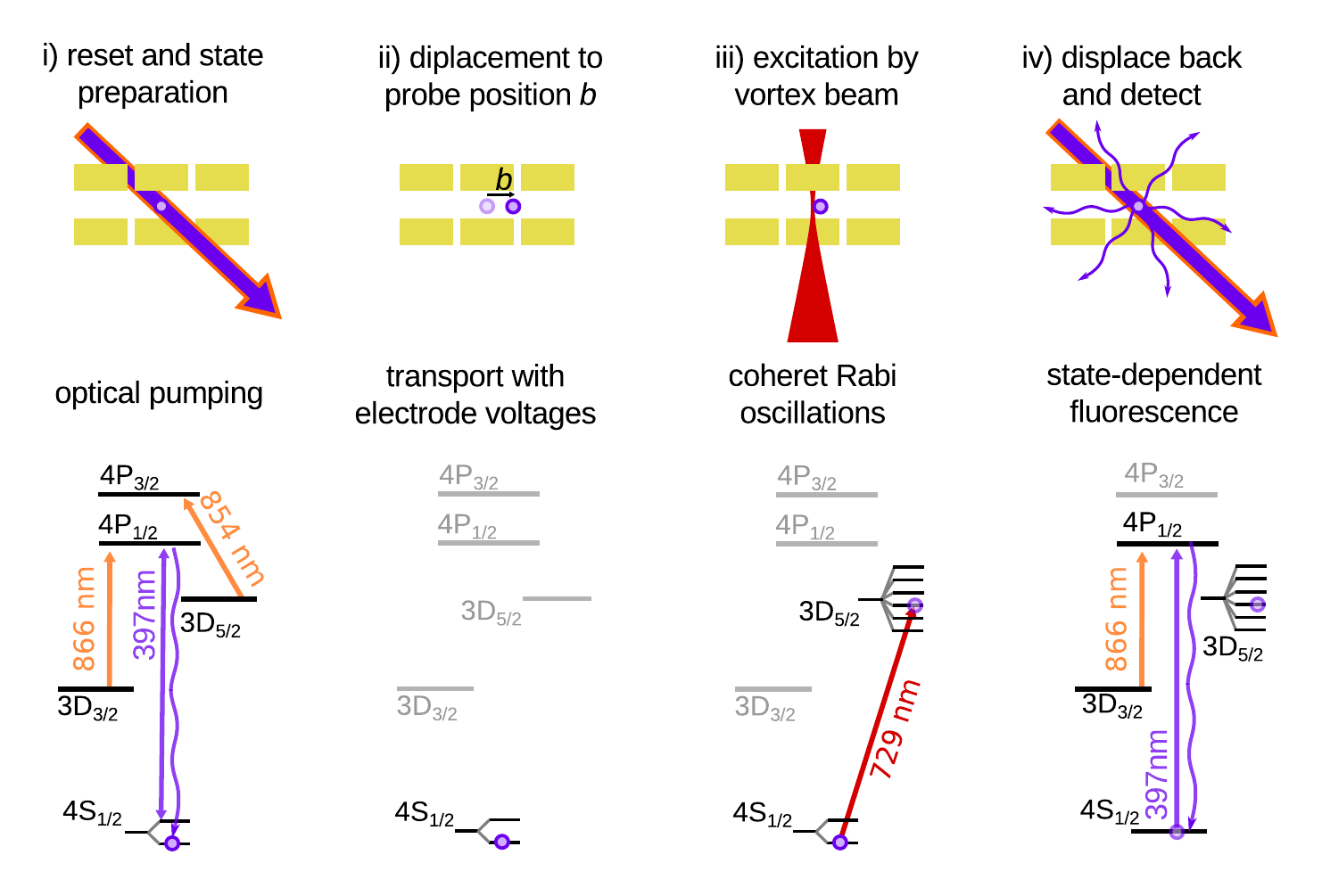}
\caption{Experimental sequence. i) First the initial state $4^2S_{1/2}$, $m_j=\pm1/2$ is prepared by optical pumping on the 397~nm transition, additionally re-pumping and state reset from lower-lying $D$ states is performed by two lasers tuned to the 866 and 854~nm transitions. ii) Next the ion is shuttled to the a given position by sweeping the electrode voltages where the interaction along the beam will be measured. iii) Following the probe beam is turned on and the ion oscillates coherently between frequency selected Zeeman sub-levels. iv) Finally the ion is shuttled back to the initial position the electronic state is read out by state dependent fluorescence on the 397~nm transition with re-pumping on the 866~nm transition. See text for more details.
}
\label{fig:experiment}
\end{figure}

\begin{figure}[h]
\centering
\includegraphics[width = 0.48 \columnwidth]{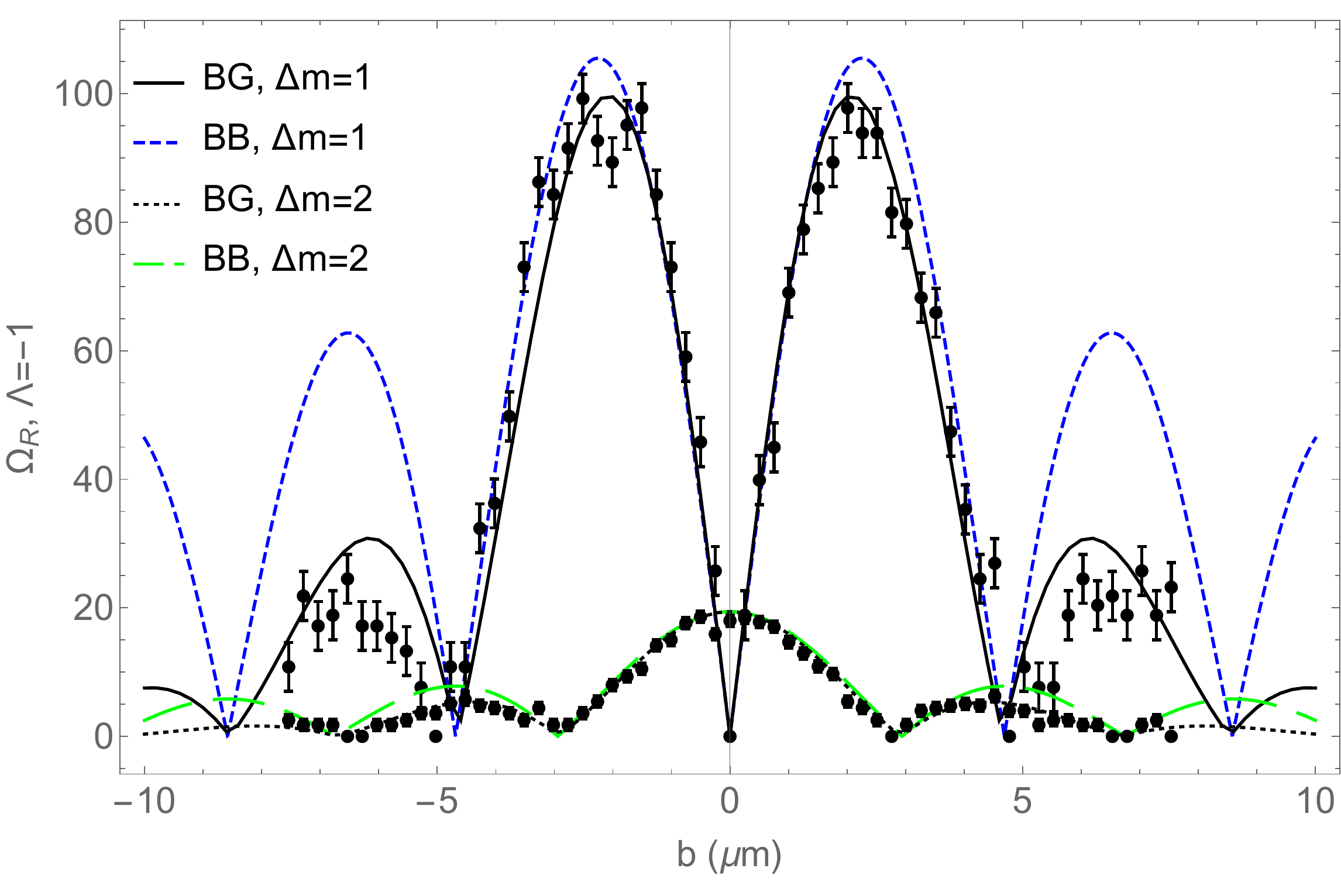}	\hfil
\includegraphics[width = 0.48 \columnwidth]{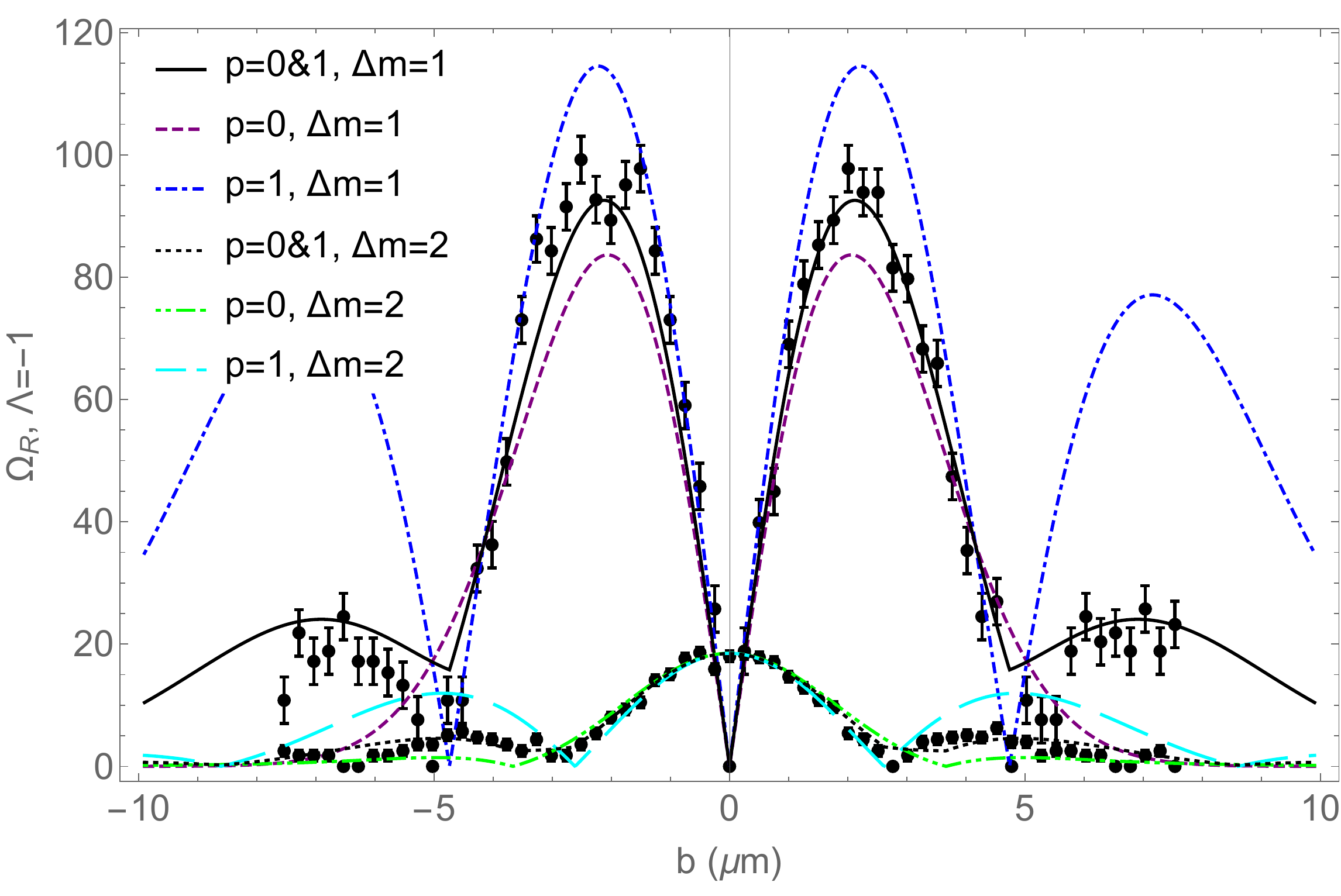}
\caption{(a) Normalized Rabi frequencies for OAM light with $m_\gamma=-2$ described with BB mode (blue solid and green long-dashed) and BG mode (black dashed and purple dotted) as a function of the impact parameter $ b$ for the transitions $\Delta m = -2$ and $\Delta m = -1$ from the ground state with $m_i=-1/2$, $\Lambda$=-1; (b)~same transitions for LG modes with parameters $p$=0, $p$=1, and their linear combination (see text for details). }
\label{fig:BB_BG_m12}
\end{figure}

\begin{figure}[h]
\centering
\includegraphics[width = 0.48 \columnwidth]{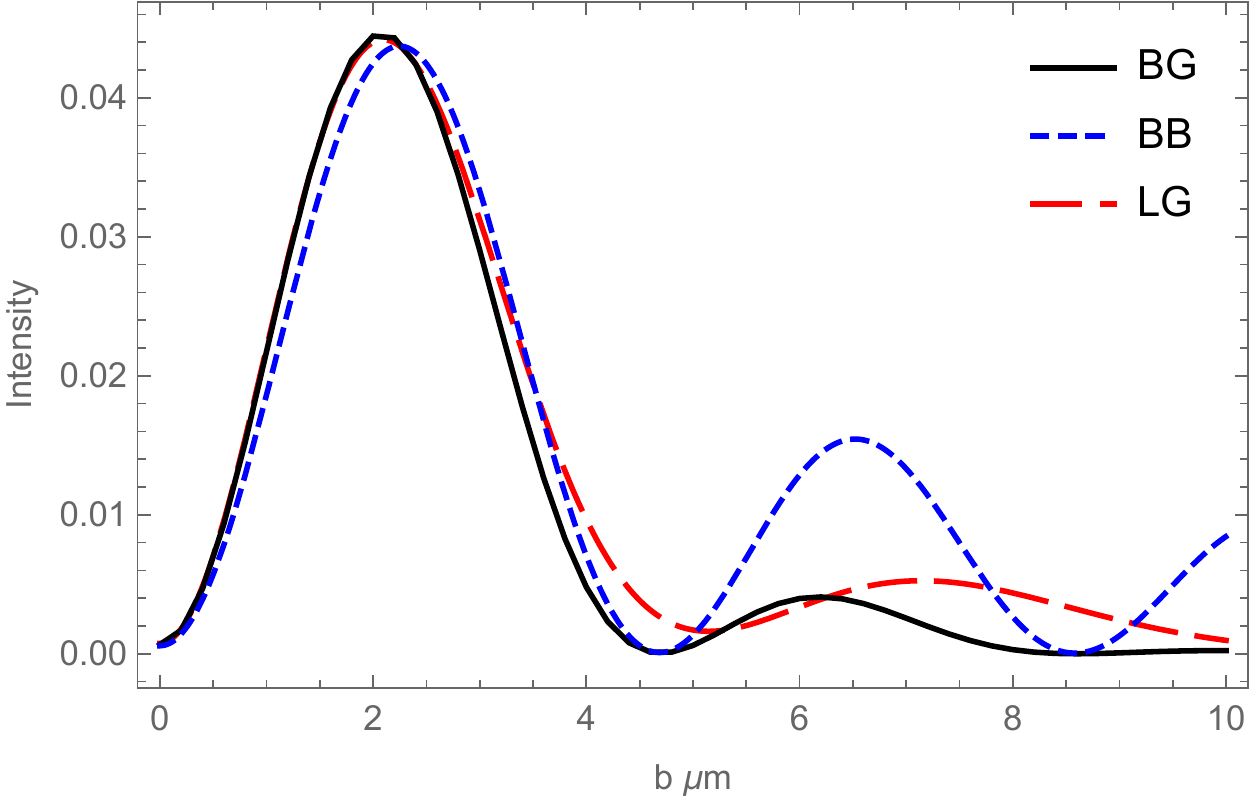}		\hfil
\includegraphics[width = 0.48 \columnwidth]{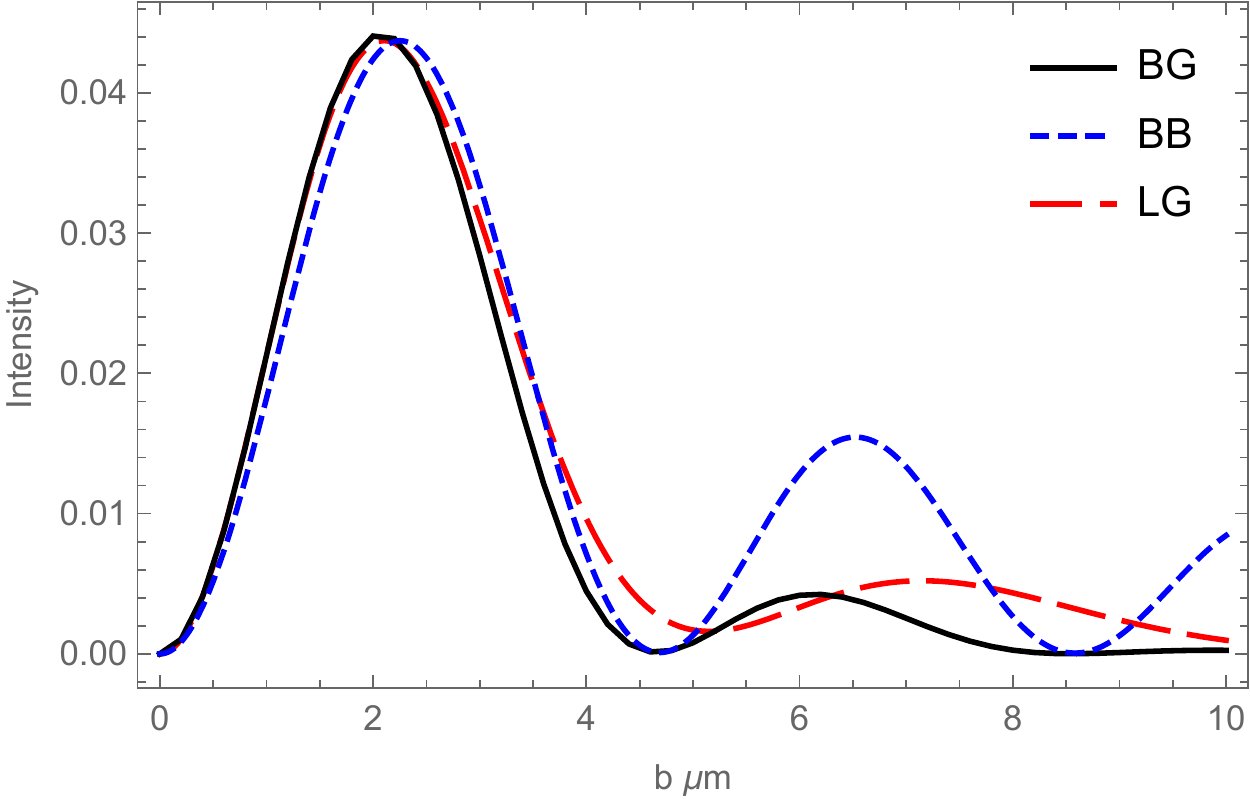}
\caption{Beam's local energy flux used in the theory fits as a function of impact parameter for BB (black solid), BG (red dotted) and LG (blue dashed) modes. Left plot: $m_\gamma=0$, right plot: $m_\gamma=2$. }
\label{fig:flux}
\end{figure}



\begin{figure}[h]
\vspace{2mm}
\centering
\includegraphics[width = \columnwidth]{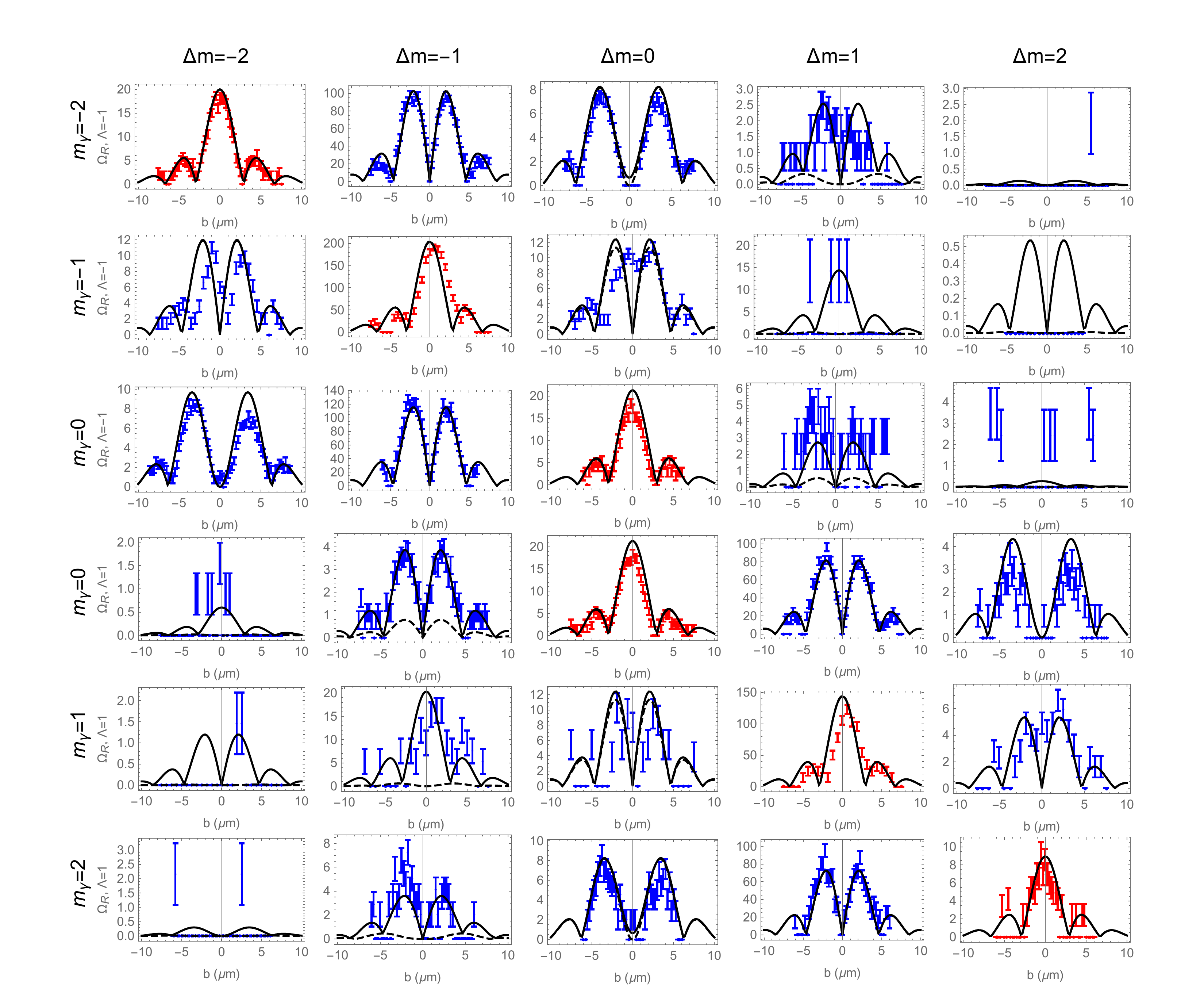}
\caption{Normalized Rabi frequencies as a function of impact parameter $b$, compared with theory predictions for BG modes (solid lines). The projection of initial atomic spin is $s_z=-1/2$. Black dashed curves correspond to the theory predictions not accounting for the opposite-sign circular polarization admixture; solid curves in green are the theory predictions with 3\% opposite-polarization admixture (by amplitude) for rows 1,3,4 and 6, and 10\% admixture for rows 2 and 5. Columns 1 through 5 correspond to the change of magnetic quantum number by $\Delta m=-2,-1,0,1,2$, respectively. Rows 1 through 6 correspond photon's angular momentum projections  $m_\gamma=-2,-1,0 (\Lambda=-1),0 (\Lambda=1),1,2$.}
\label{fig:fullmBG}
\end{figure}

\begin{figure}[h]
\vspace{2mm}
\centering
\includegraphics[width = \columnwidth]{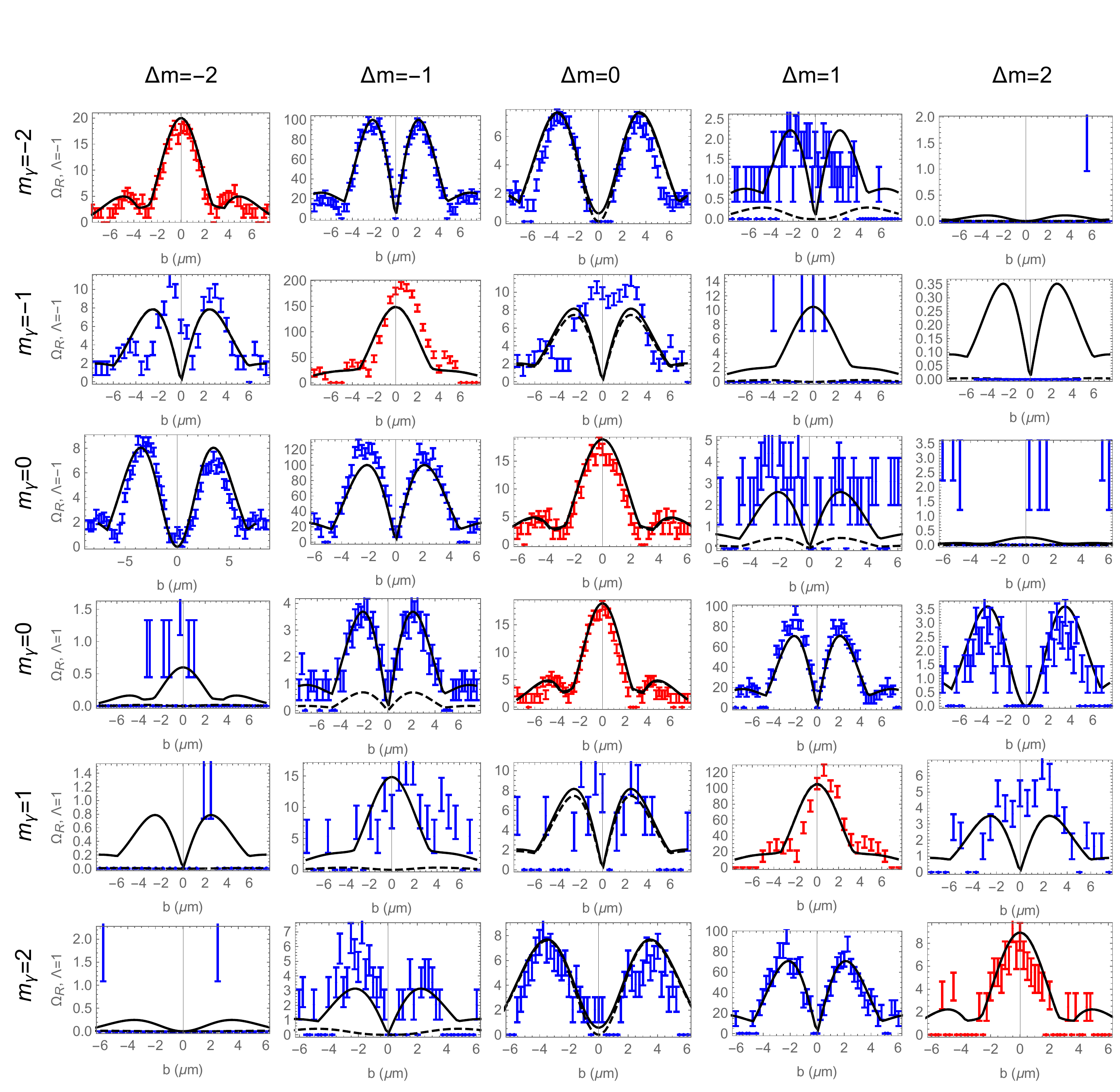}
\caption{Same as Fig.\ref{fig:fullmBG}, described with LG modes.}
\label{fig:fullmLG}
\end{figure}

\begin{figure}[h]
\vspace{2mm}
\centering
\includegraphics[width = \columnwidth]{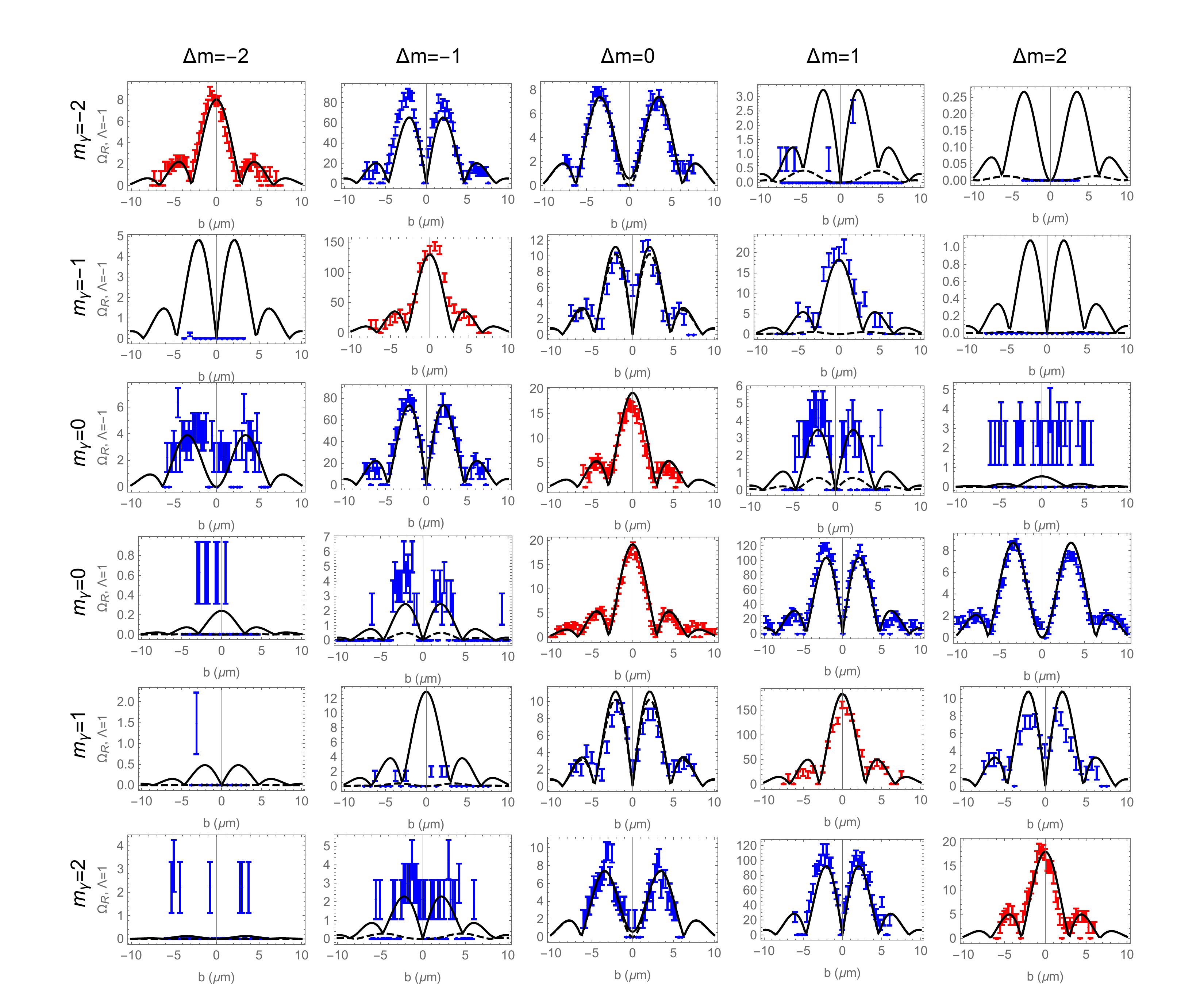}
\caption{Same as Fig.\ref{fig:fullmBG} (BG modes), but for the initial atomic state with an opposite $s_z$=1/2.}
\label{fig:fullpBG}
\end{figure}

\begin{figure}[h]
\vspace{2mm}
\centering
\includegraphics[width = \columnwidth]{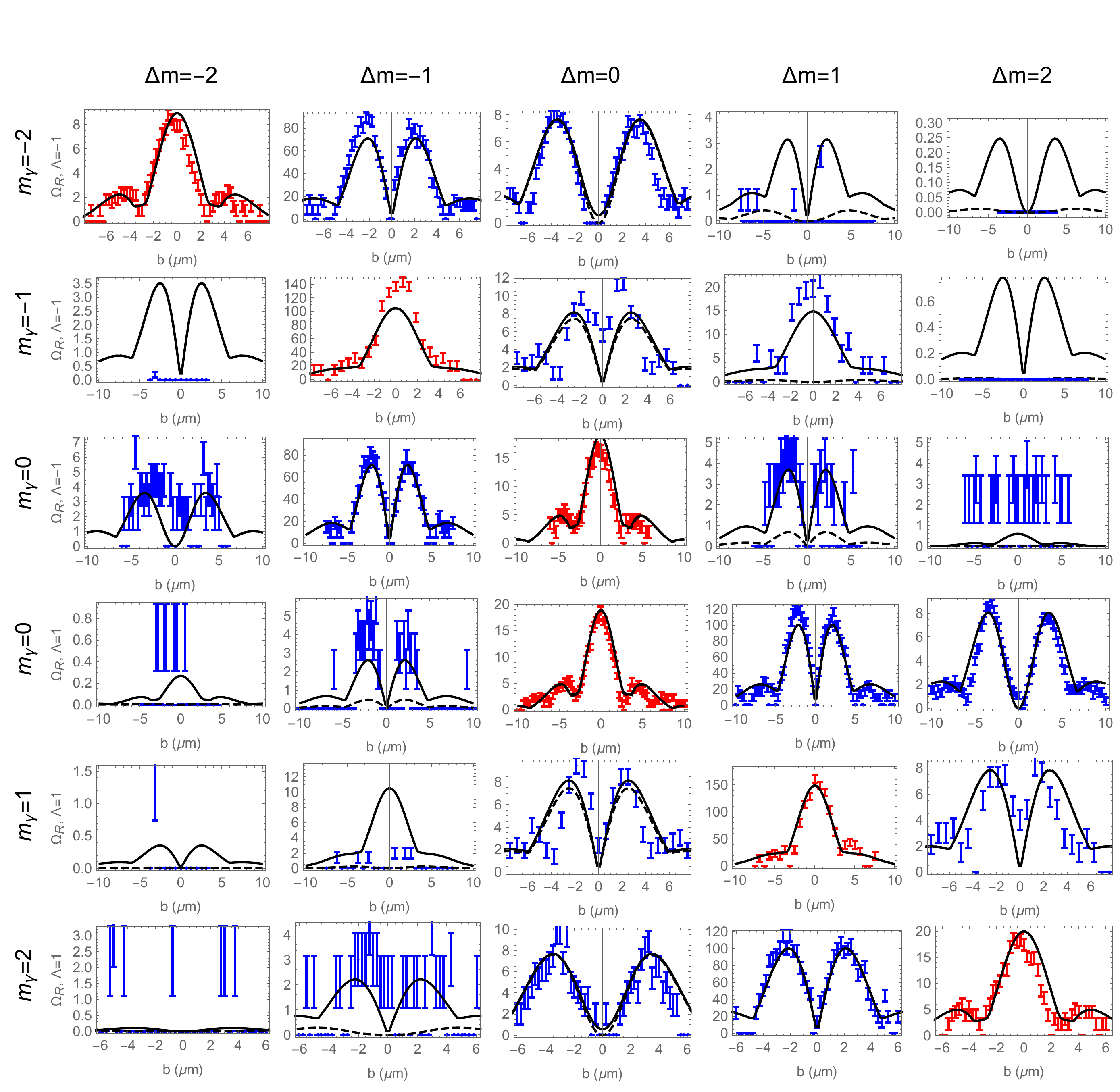}
\caption{Same as Fig.\ref{fig:fullpBG} ($s_z=1/2$), theory with LG modes.}
\label{fig:fullpLG}
\end{figure}


\begin{figure}[h]
\vspace{2mm}
\centering
\includegraphics[width = \columnwidth]{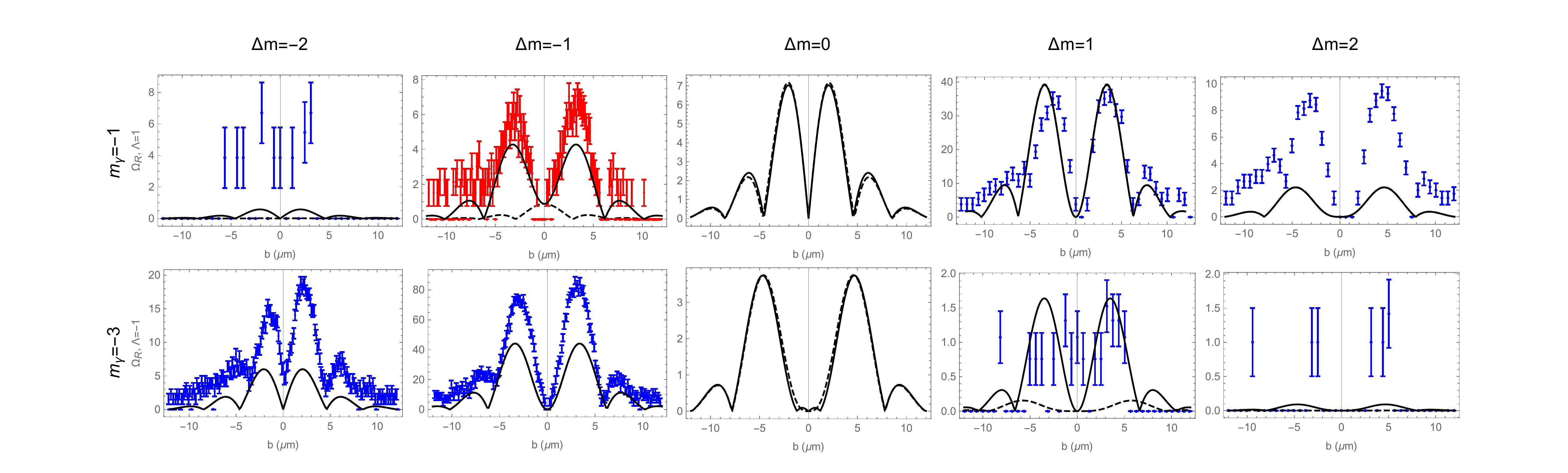}
\caption{Similar to Fig.~\ref{fig:fullmBG}, but here the photons have $\left| \overline m_\gamma \right| = 2$.   (In the paraxial limit, $\overline m_\gamma$ would be the orbital angular momentum.)  The quantum numbers are  $m_\gamma=-1, \Lambda=1$ (upper plots) and $m_\gamma=-3, \Lambda=-1$ (lower plots), $s_z=-1/2$. The curves are the Bessel-Gauss theory curves, with the dashed curves having no admixture of opposite helicity photons, and the solid curves have 10\% by amplitude of opposite helicity photons.}
\label{fig:OAM2_BG}
\end{figure}

\begin{figure}[h]
\vspace{2mm}
\centering
\includegraphics[width = \columnwidth]{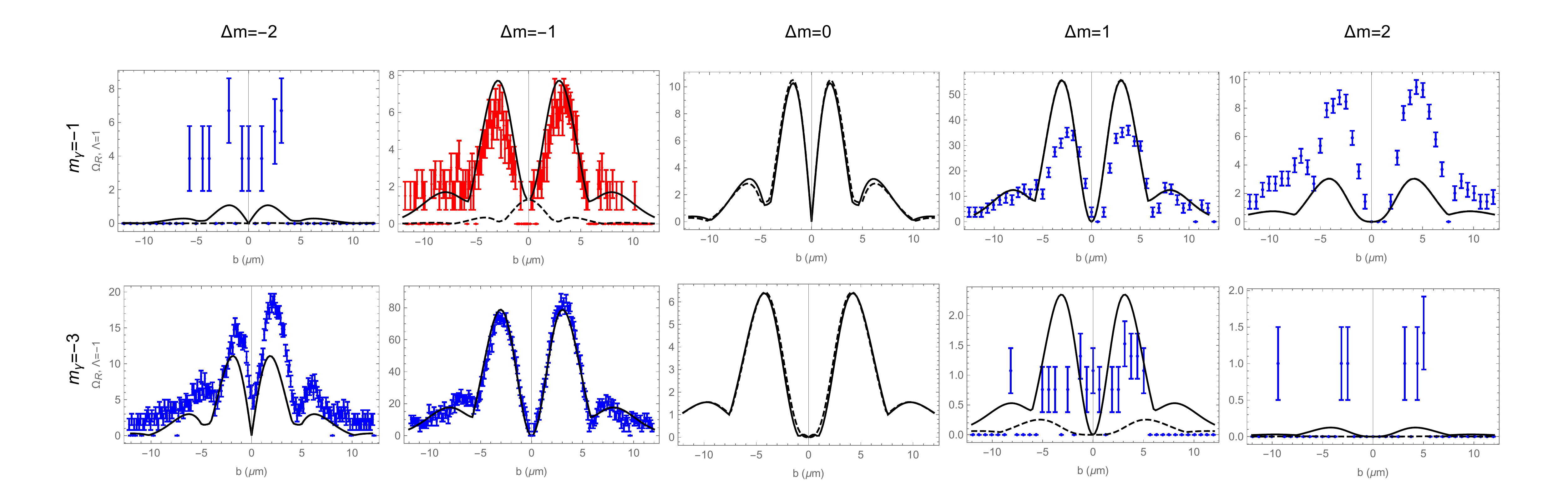}
\caption{Same as Fig~\ref{fig:OAM2_BG}, but with theory predictions for LG modes.}
\label{fig:OAM2_LG}
\end{figure}

\begin{figure}[h]
\centering
\includegraphics[scale=0.7]{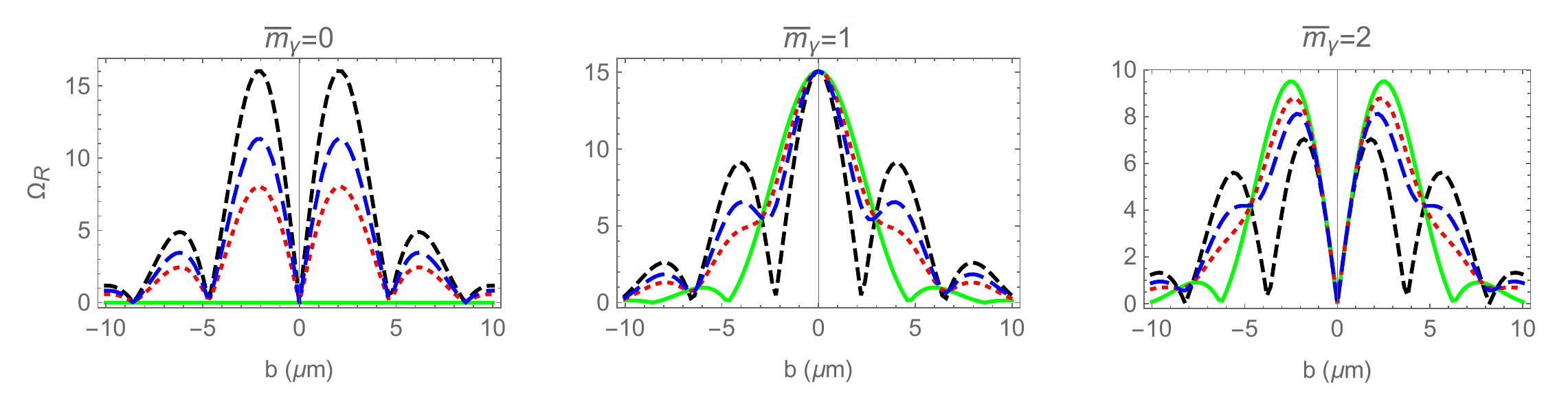}
\includegraphics[scale=0.45]{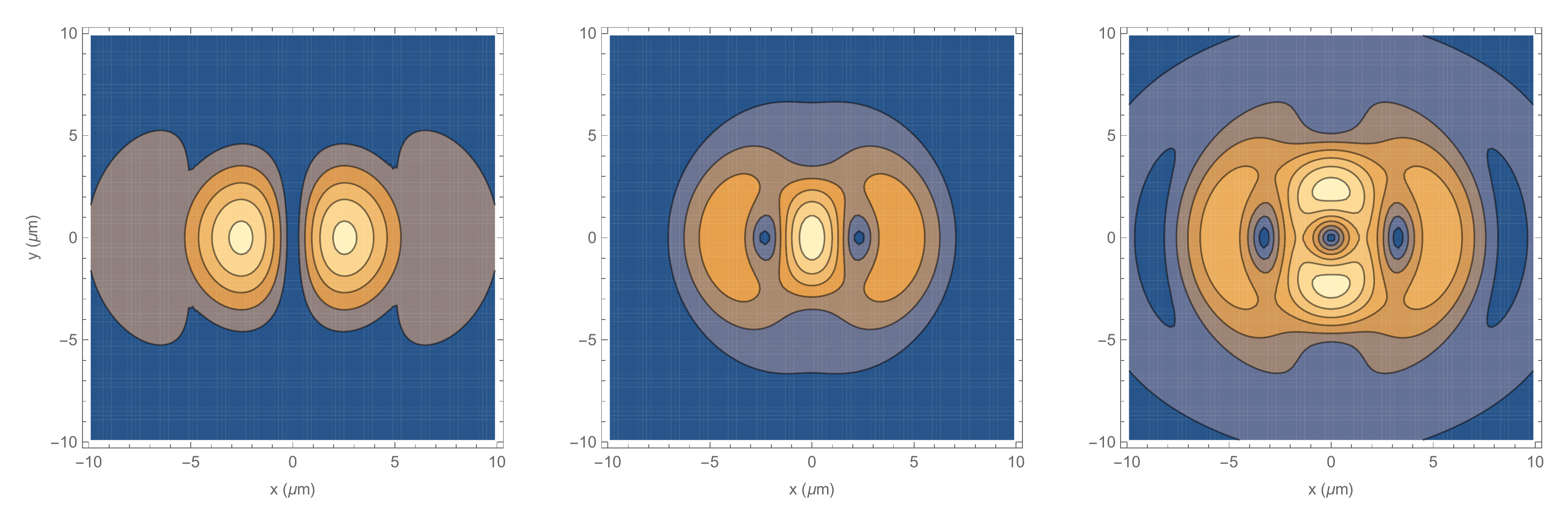}
\caption{Predictions, using BG modes, for the azimuthal dependence of $\Delta m=0$ transition amplitudes when there is a linearly polarized OAM beam. In the upper row we have $\overline m_\gamma=0, 1$, and $2$, as labeled, each plot with four azimuthal angles:
$\phi_b$=0 (dashed black), $\pi/2$ (solid green), $\pi/3$ (red dotted), and $\pi/4$ (blue long-dashed).  In the lower row are contour plots showing the magnitude of the amplitude at each location  in the $x$-$y$ plane.  Lighter shades (white and yellow) indicate larger amplitudes and darker shades (blue) indicate smaller amplitudes. }
\label{fig:azimuthal}
\end{figure}


\end{document}